\documentclass[reprint,amsmath,amssymb,aps,pre,nofootinbib,floatfix]{revtex4-2}

\usepackage{graphicx}
\usepackage{bm}
\usepackage{tikz}
\usetikzlibrary{arrows.meta,calc}
\usepackage{xcolor}
\usepackage{hyperref}
\hypersetup{
    colorlinks=true,
    linkcolor=blue,
    citecolor=blue,
    urlcolor=blue
}

\newcommand{\E}[1]{\mathbb E\!\left[#1\right]}
\newcommand{\Pp}{\mathbb P}
\newcommand{\Var}{\operatorname{Var}}
\newcommand{\Cov}{\operatorname{Cov}}
\newcommand{\Tr}{\operatorname{Tr}}
\newcommand{\sign}{\operatorname{sgn}}
\newcommand{\1}{\mathbf1}
\newcommand{\ee}{\hat{\mathbf e}}
\newcommand{\dd}{\mathrm d}
\newcommand{\e}{\mathrm e}
\newcommand{\ms}{\mu_{{\rm sig},\ell}}
\newcommand{\qp}{q_\perp}
\newcommand{\lt}{\widetilde\lambda}
\newcommand{\mb}{\bar m}

\DeclareMathOperator*{\argmin}{arg\,min}
\DeclareMathOperator*{\extr}{extr}

\begin{document}

\title{Statistical mechanics of linear classification on hierarchical tree data}

\author{J\'er\^ome Garnier-Brun}
\email[Corresponding author: ]{jerome.garnier@unibocconi.it}
\author{Federica Kulka}
\author{Carlo Lucibello}
\affiliation{%
Bocconi Institute for Data Science and Analytics (BIDSA) \\ Department of Computing Sciences, Bocconi University, Milan, Italy
}

\date{\today}

\begin{abstract}
Deep networks owe much of their success to their ability to exploit the structure in data, from the grammar of language to the multi-scale organization of images. Yet, many statistical mechanics approaches to structured data rely on Gaussian equivalence, whose assumptions can fail when a few latent variables influence all input coordinates. In this work, we take a step beyond this setting with a simple setup, in which a linear classifier is trained to infer the root of the classical broadcast process on a tree from its $N$ leaves. Despite this simplicity, the data already display non-Gaussian statistics along the direction carrying the label, as well as strong finite-size effects. By retaining the exact law along this direction and approximating the transverse fluctuations as Gaussian, we obtain a finite-depth replica theory that accurately describes the measured learning curves. At large depth and sufficiently weak regularization, we uncover a double descent whose two stages are separated in scale. In a range of broadcast noise below the reconstruction threshold, the first descent occurs at a number of samples scaling as $N^\omega$, with $0<\omega<1$ set by the data statistics, whereas the overfitting peak occurs at order $N$ samples. Beyond linear classifiers, this setup could serve as a controlled starting point to study deeper architectures and generative models on hierarchical data.
\end{abstract}

\maketitle

\section{Introduction}

Statistical mechanics has long provided tools for studying artificial neural networks, from the storage capacity of the perceptron \cite{gardner1988space,gardner1988optimal} to the learning curves of teacher--student scenarios \cite{gardner1989three,seung1992statistical} (see \cite{Engel_VandenBroeck_2001,gabrie2023neural} and references therein), and it remains a tool of choice to describe learning in high dimension, including for deep networks \cite{barbier2026statistical} and attention-based models \cite{erba2025bilinear,boncoraglio2026bayes}. There is, however, a broad consensus that learning from real data relies on exploiting its structure, which is precisely where the statistical mechanics theory currently lags behind \cite{mezard2024spin}. Indeed, the core of the statistical mechanics approach rests on concentration of measure, which is natural for inputs with many independent degrees of freedom but a priori at odds with genuinely low-dimensional structure. In the latter, the statistics of the inputs are organized across a number of scales much smaller than the dimension $N$ of the data, e.g.\ as few as $O(\log N)$, with a handful of latent variables at the top of hidden hierarchies influencing all the inputs. On the one hand, the tractable models of structured data have therefore remained relatively tame; in the hidden manifold model of \citet{goldt2020modeling}, for instance, the latent dimension still grows proportionally to $N$. On the other hand, models with a genuinely hierarchical structure, such as the random hierarchy model (RHM) of \citet{cagnetta2024how} and related tree-based models~\cite{garnier2024transformers}, have so far escaped asymptotic analyses of learning. Their study has instead relied on signal-to-noise arguments and comparisons with controlled oracles, or on proofs that suitable architectures can learn them efficiently~\cite{mei2024unets,ren2026provable,gaitonde2026hierarchical}. Other theoretical works on hierarchical learning place the hierarchy in the target function rather than in the data, the inputs remaining Gaussian~\cite{wang2023learning,fu2024learning,dandi2025computational,defilippis2026optimal,tabanelli2026deep,wortsman2026scaling,dandi2026neural}.

In this work, we take a first step towards a theory of learning from genuinely hierarchical data by studying the empirical risk minimization of a linear classifier tasked with inferring the root of a tree-based broadcast process from its leaves. In our setting, all variables are binary spins, and the process has a single parameter that is the probability that a child differs from its parent (see Fig.~\ref{fig:tree}), a drastic simplification compared with the vocabularies and production rules of the RHM. Despite this simplicity, it already displays key technical challenges that stem directly from the hierarchical nature of the data: (i) non-Gaussian statistics along the direction carrying the label, for which Gaussian equivalence fails; (ii) strong finite-size effects; (iii) a sample complexity set by the structure of the data. By isolating the non-Gaussian direction and keeping track of the finite depth in a replica computation, we account for all three, and describe the learning curve down to sublinear numbers of samples. In the limit of large depth, this allows us to uncover a double descent whose stages are separated in scale, with a first descent set by the statistics of the data alone at a number of samples $\sim N^\omega$, $\omega<1$, followed, at sufficiently weak regularization, by an overfitting peak and second descent in the proportional regime.

Compared to the RHM, for which shallow networks suffer from the curse of dimensionality~\cite{cagnetta2024how}, the broadcast model offers an interesting opportunity, as a simple linear statistic of the leaves already achieves a classification error close to the Bayes error, so that the problem is meaningful already at the level of the linear classifier we study here. Belief propagation, the Bayes-optimal estimator, can nonetheless improve on the census through a hierarchical, nonlinear processing of the leaves. The same model therefore opens the way to studying more expressive architectures beyond the present work, for which the approach developed here provides a starting point.

The paper is organized as follows. In Sec.~\ref{sec:data}, we present the broadcast model and derive the statistics relevant to linear learning, namely the hierarchical structure of the leaf covariance, the non-Gaussian law of the census and its finite-depth corrections. In Sec.~\ref{sec:theory}, we develop the replica theory of empirical risk minimization, first in the proportional regime at infinite depth (Sec.~\ref{sec:proportional}), then at finite depth (Sec.~\ref{sec:finite-size}), and finally for a sub-proportional number of samples (Sec.~\ref{sec:sample_complexity}), where the scale-separated double descent emerges. We discuss our results and possible extensions in Sec.~\ref{sec:discussion}.

\section{Tree-based hierarchical data}
\label{sec:data}

\subsection{Presentation of the model}

\begin{figure}
\centering
\begin{tikzpicture}[
  x=0.7cm, y=0.85cm,
  spin/.style={circle,draw,line width=0.6pt,minimum size=2.4mm,inner sep=0pt},
  up/.style={spin,fill=black},
  down/.style={spin,fill=white},
  every path/.style={line width=0.7pt},
]
  \node[up] (r) at (3.5,0) {};
  \node[up] (a) at (1.5,-1) {};
  \node[down] (b) at (5.5,-1) {};
  \node[up]   (aa) at (0.5,-2) {};
  \node[down] (ab) at (2.5,-2) {};
  \node[down] (ba) at (4.5,-2) {};
  \node[up]   (bb) at (6.5,-2) {};
  \foreach \i/\s in {0/up,1/up,2/down,3/up,4/down,5/down,6/up,7/down}
    \node[\s] (l\i) at (\i,-3) {};
  \draw (r) -- node[above left=-1pt] {$1-\varepsilon$} (a);
  \draw (r) -- node[above right=-1pt] {$\varepsilon$} (b);
  \foreach \u/\v in {a/aa,a/ab,b/ba,b/bb,aa/l0,aa/l1,ab/l2,ab/l3,ba/l4,ba/l5,bb/l6,bb/l7}
    \draw (\u) -- (\v);
  \node[above=1pt] at (r.north) {$x_0$};
  \draw[gray,dashed,line width=1.1pt,rounded corners=3pt]
    ([shift={(-3pt,-3pt)}]l0.south west) rectangle ([shift={(3pt,3pt)}]l7.north east);
  \node[below=4pt] at ($(l0.south)!0.5!(l7.south)$) {$\{x_i\}_{i=1}^{N}$};
  \draw[line width=0.4pt,{Stealth[length=4pt]}-{Stealth[length=4pt]}] (8.2,0) -- (8.2,-3)
    node[midway,right] {$\ell=\log_2 N$};
\end{tikzpicture}
\caption{Broadcast tree model. The root spin $x_0\in\{-1,+1\}$ is uniform; each edge independently flips its parent spin with probability $\varepsilon$. The $N=2^\ell$ leaves $\{x_i\}_{i=1}^{N}$ of the depth-$\ell$ tree form the input. Shown is one realization with $\ell=3$ (black: $+1$, white: $-1$).}
\label{fig:tree}
\end{figure}
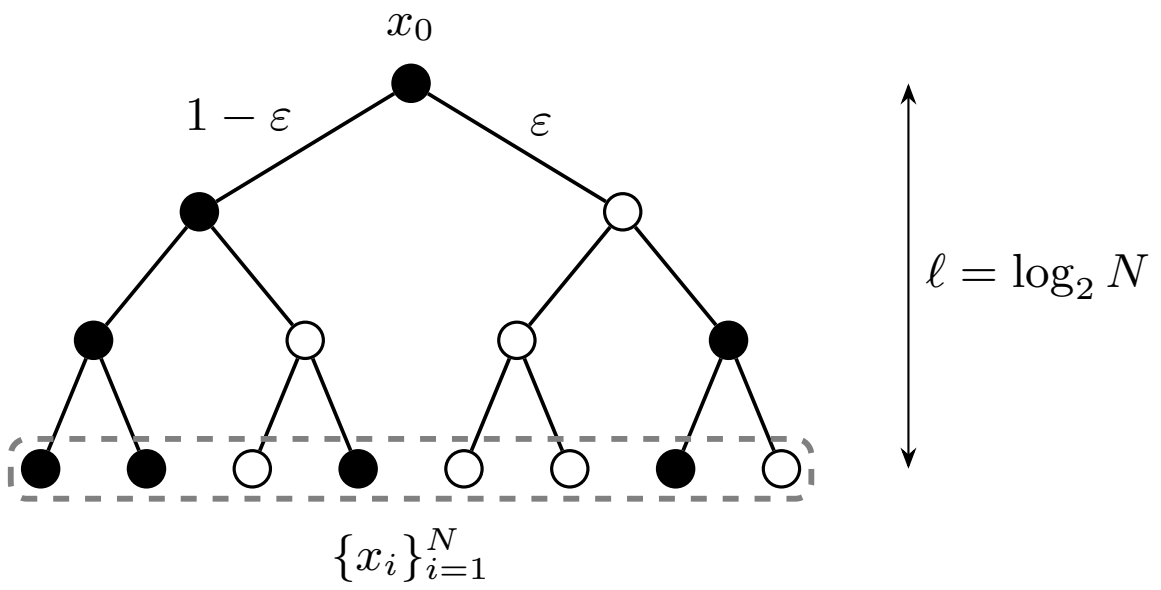

We consider the broadcast process on a binary tree of depth $\ell$, illustrated in Fig.~\ref{fig:tree}. The root spin $x_0=\pm1$ is drawn uniformly, $\Pp(x_0 = 1) = 1/2$, and each node passes its value to its two children, which independently copy it with probability $1-\varepsilon$ and flip it otherwise. Each edge thus acts as a binary symmetric channel with flip probability $\varepsilon$, which plays the role of a noise. We restrict to the ``ferromagnetic'' case $0\leq\varepsilon\leq\frac{1}{2}$, the ``anti-ferromagnetic'' one $\frac{1}{2}\leq\varepsilon\leq1$ following by symmetry. Repeating this Markovian step over $\ell$ generations produces the $N = 2^\ell$ leaf spins $\{x_i\}_{i=1,\dots,N}$, which form the input data. This process has a long history as a canonical model of information flow on trees, in probability and information theory as well as in statistical physics, where it can be understood as an Ising model on a tree~\cite{bleher1995purity,evans2000broadcasting,mossel2003information,mossel2004survey,mezard2006reconstruction,mezard2009information}.

For subsequent calculations, it is convenient to describe the broadcasting process with the transition matrix,
\begin{equation}
    \bm{\pi}(\varepsilon) = 
    \begin{bmatrix}
        1-\varepsilon & \varepsilon\\
        \varepsilon & 1-\varepsilon
    \end{bmatrix},
\end{equation}
whose eigenvalues are $1$ and $\theta$, where we have introduced the complementary quantity
\begin{equation}
    \theta = 1 - 2\varepsilon, \qquad 0\leq \theta \leq 1,
\end{equation}
which can be understood as the parent-child correlation since $\E{x_{\rm child} \mid x_{\rm parent}} = \theta x_{\rm parent}$.

In this work, we study linear classifiers predicting the root from the leaves $\bm x=(x_1,\dots,x_N)$, of the form $\hat x_0 = \sign(\bm w\cdot\bm x)$, with $\bm w$ learned from a finite number $P$ of examples. A natural reference point is then the \textit{census}, defined simply as
\begin{equation}
    C^{(\ell)} = \sum_{i=1}^N x_i, \qquad \hat{x}_0^{\mathrm{census}} = \sign C^{(\ell)},
\end{equation}
i.e.\ a majority rule on the leaves, $\bm w = \mathbf{1}$. Throughout, a zero classification margin $\bm w \cdot \bm x$ is resolved by a fair coin.

For the $\pm 1$ symmetric channel, it is known~\cite{mossel2003information,mossel2004survey} that the census estimator achieves non-trivial reconstruction accuracy when $\varepsilon$ is sufficiently small for the generative process to carry some detectable signal despite the exponentially increasing amount of fluctuations induced by going down the tree. Asymptotically, this occurs through a phase transition at the critical value $\varepsilon_c$ defined by the celebrated Kesten--Stigum (KS) criterion~\cite{kesten1966limit}, which for our binary tree reads
\begin{equation}
    r = 2\theta^2 > 1,
\end{equation}
i.e.\ $\varepsilon < \varepsilon_c = \frac{1}{2} (1 - 1/\sqrt{2}) \approx 0.146$. We refer to $r$ as the KS parameter. The KS criterion extends to a regular tree of branching ratio $d$, with $r=d\theta^2$ and $N=d^\ell$. The analysis can be adapted to that setting, but the spectral multiplicities and other numerical coefficients also depend on $d$; here we restrict to binary trees. Changing instead the number of states of each variable, e.g.\ to a $q$-state Potts broadcast, would lead to a much more complex setting, in which reconstruction becomes possible below the KS threshold for large enough $q$~\cite{mezard2006reconstruction}.

As mentioned in the introduction, the census is not generally Bayes-optimal. Since the generative model is a tree, the optimal estimator is instead computed exactly by belief propagation (BP)~\cite{mezard2009information}, which reduces here to a simple upward recursion on the leaf spins (Appendix~\ref{app:bp}). In the binary symmetric channel we consider, the census nonetheless loses relatively little compared to this baseline, as both estimators become uninformative at the same value $\varepsilon = \varepsilon_c$~\cite{bleher1995purity,evans2000broadcasting}, and for $\varepsilon<\varepsilon_c$ the accuracy of the census remains close to that of BP, see Fig.~\ref{fig:bp-census} of Appendix~\ref{app:bp}.

\subsection{Exact statistics}

We start by describing the statistics of the broadcast data. As the task that we consider in this work is to recover the root spin given the full sequence of leaves, we focus on root-conditional statistics.

\subsubsection{Leaf correlation structure}
\label{sec:correls}

As stated above, along a root-to-leaf path the transition matrix $\bm \pi$ has eigenvalues $1$ and $\theta$. Hence, writing $x_0 = y$,
\begin{equation}
 \Pp(x_i=\sigma\mid y)=\tfrac{1}{2}(1+\sigma y\theta^\ell),
 \qquad \E{x_i\mid y}=y\theta^\ell.
\end{equation}
Beyond these individual leaf-root correlations, a key element of this data model is that leaves are correlated even at long range, as they inherit the fluctuations of their common ancestors in the hierarchy. For two leaves $(i,j)$ with lowest common ancestor at depth $k_{ij}$ from the root, conditional independence below that ancestor gives
\begin{equation}
 \E{x_ix_j\mid y}=\theta^{2(\ell-k_{ij})},
\end{equation}
and thus 
\begin{equation}
     \Sigma_{ij}:=\Cov(x_i,x_j\mid y) =\theta^{2(\ell-k_{ij})}-\theta^{2\ell}.
 \label{eq:covariance}
\end{equation}
Taking $k_{ii}=\ell$ gives the diagonal, $\Sigma_{ii}=1-\theta^{2\ell}$, the variance of a leaf given the root. Covariance naturally increases with the common ancestor's depth, i.e.\ as the two leaves get closer in the tree, while the two halves of the tree are independent conditional on the root. Covariance entries alone do not reveal which collective fluctuations are large, notably those affecting the census. Instead, we diagonalize $\bm \Sigma$ recursively in the depth, using the self-similarity of the tree.

Since the two halves of the tree are independent given the root, ordering the leaves by their tree positions makes the covariance of a depth-$\ell$ tree block diagonal,
\begin{equation}
 \bm \Sigma^{(\ell)}=
 \begin{bmatrix}
  \mathbf{A}_{\ell-1} & 0\\
  0 & \mathbf{A}_{\ell-1}
 \end{bmatrix},
\end{equation}
where $\mathbf{A}_{\ell-1}$ is the covariance of the leaves of one half-tree, conditioned on the global root. For two such leaves, the common ancestor sits one level closer to the top of their half-tree than in the full tree, and Eq.~\eqref{eq:covariance} gives
\begin{equation}
 \mathbf{A}_{\ell-1}=\bm{\Sigma}^{(\ell-1)}+(1-\theta^2)\,\theta^{2(\ell-1)}\,\1\1^\top.
 \label{eq:cov-recursion}
\end{equation}
The rank-one term is the variance of the half-tree's own root given the global root, which is transmitted to all its leaves with amplitude $\theta^{\ell-1}$.

All rows of $\bm \Sigma^{(\ell-1)}$ contain the same entries up to a permutation, so they have equal sums and the uniform vector $\1$ is an eigenvector. Since $\1\1^\top$ acts only along $\1$, with eigenvalue $\1^\top\1=2^{\ell-1}$, the update \eqref{eq:cov-recursion} leaves all eigenvectors orthogonal to $\1$ and their eigenvalues unchanged, and shifts the uniform eigenvalue by
\begin{equation}
 2^{\ell-1}(1-\theta^2)\,\theta^{2(\ell-1)}=(1-\theta^2)\,r^{\ell-1},
\end{equation}
with $r=2\theta^2$ the KS parameter given above.

Each eigenvector of $\mathbf{A}_{\ell-1}$, placed on either block, then gives an eigenvector of $\bm \Sigma^{(\ell)}$, so all multiplicities double. In particular, the two copies of the shifted uniform vector combine into the uniform vector of the full tree and the root contrast $\1_L-\1_R$ ($\1_{L,R}$ being the indicators of the two halves), which share the same eigenvalue. Starting from a single layer, $\bm \Sigma^{(1)}=(1-\theta^2)\mathbf{I}_2$, the uniform eigenvalue of a depth-$k$ tree obeys $\mu_k=\mu_{k-1}+(1-\theta^2)r^{k-1}$, with $\mu_1=1-\theta^2$. Hence, for $0<\theta<1$, $\bm \Sigma^{(\ell)}$ has $\ell$ distinct eigenvalues
\begin{equation}
 \mu_k =(1-\theta^2)\sum_{j=0}^{k-1}r^j = (1-\theta^2)\,\dfrac{r^k-1}{r-1}, \quad (r\neq1),
 \label{eq:tree-eigenvalues}
\end{equation}
for $1\le k\le\ell$, while precisely at the KS transition $r=1$ we have $\mu_k = k/2$. Their multiplicities are $2^{\ell-k}$ for $k<\ell$, and two for the largest one, $\mu_\ell$. Note that $\mu_k$ does not depend on $\ell$, increasing the depth instead only adds a new, larger eigenvalue and changes the multiplicities.

Importantly, the census direction is an eigenvector,
\begin{equation}
 \bm \Sigma\ee=\ms\ee,\qquad \ms=\mu_\ell,\qquad \ee=\frac{\1}{\sqrt N},
\end{equation}
so that the census and transverse fluctuations are uncorrelated. Beyond the census direction, only the spectrum of $\bm \Sigma$, not its eigenvectors, will enter the learning theory. As we will see, it is convenient to separate the signal direction from the orthogonal (transverse) directions. We define the full and transverse spectral measures by
\begin{equation}
\begin{aligned}
 \rho_\ell&=\sum_{k<\ell}2^{-k}\delta_{\mu_k}+2^{1-\ell}\delta_{\mu_\ell},\\
 \rho_{\perp,\ell}&=\rho_\ell-N^{-1}\delta_{\ms}.
\end{aligned}
\label{eq:measures}
\end{equation}
The transverse measure removes one top mode and has mass $1-1/N$; its averages retain normalization by $N$. The limiting spectral density is
\begin{equation}
 \rho(\mu)=\sum_{k=1}^\infty2^{-k}\delta(\mu - \mu_k).
 \label{eq:rho-limit}
\end{equation}
The eigenvalues remain bounded for $r<1$, whereas for $r>1$, $\mu_k\asymp r^k$ grows without bound, but most of the mass, carried by small $k$, remains at order-one eigenvalues. As a result, for fixed $1<r<2$ the spectral moments $\langle\mu^p\rangle_\rho=\sum_k2^{-k}\mu_k^p$ of the limiting density diverge when $r^p\ge2$. In particular, the second moment diverges for $\sqrt2\le r<2$, whereas the first moment, which is the variance of a leaf given the root, simply tends to one. Bounded resolvents, however, still have finite limiting averages (see below), as the diverging eigenvalues carry a vanishing mass.

\subsubsection{Non-Gaussianity of the census}
\label{sec:census}

We now focus on the census $C^{(\ell)} = \sum_i x_i$, the reference linear statistic introduced above. The census mean is $y(2\theta)^\ell$, and its conditional variance is $N\ms$. Equivalently, the projection $\ee\cdot\mathbf x$ on the unit census direction $\ee=\1/\sqrt N$ has mean $y s_\ell$ with signal amplitude $s_\ell=\sqrt N\theta^\ell$, so that $s_\ell^2=r^\ell$. For $\theta>0$, we introduce the normalized signed census, whose infinite-depth limit is nondegenerate for $1<r<2$
\begin{equation}
 c_\ell:=\frac{yC^{(\ell)}}{(2\theta)^\ell}.
\end{equation}
The endpoint $\theta=0$, where the leaves carry no information on the root, is left aside. By definition, $\E{c_\ell}=1$, while the normalized census variance is given by
\begin{equation}
 \gamma_\ell:=\Var(c_\ell)=\frac{\ms}{s_\ell^2}
 =(1-\theta^2)\sum_{j=1}^{\ell}r^{-j}.
 \label{eq:gamma-ell}
\end{equation}
For $r>1$, we may write
\begin{equation}
 \gamma_\ell=\gamma_\infty(1-r^{-\ell}),\qquad
 \gamma_\infty=\frac{1-\theta^2}{2\theta^2-1}.
 \label{eq:gamma-inf}
\end{equation}
At KS, Eq.~\eqref{eq:gamma-ell} instead gives $\gamma_\ell=\ell/2$, which diverges with the depth. The signal-to-noise ratio of the normalized census is straightforwardly $\mathrm{SNR}_\ell=1/\sqrt{\gamma_\ell}$.

The full law of $c_\ell$ can be computed exactly at finite $\ell$ by a convolution recursion over subtrees (Appendix~\ref{app:census-law}). This exact computation however becomes impractical at large depth, and importantly does not by itself give access to the $\ell\to\infty$ asymptotics. Instead, we characterize the normalized census through a recursion in distribution over the depth. Let $\eta_1,\eta_2$ be independent edge signs with $\Pp(\eta_i=-1)=\varepsilon$, so that $\E{\eta_i}=\theta$. The two children of the root carry the root spin up to these signs, and each roots an independent subtree of depth $\ell-1$. Since the census is the sum of the two subtree censuses, the normalized census obeys the recursion in distribution
\begin{equation}
 c_\ell\overset{d}{=}\frac{\eta_1c_{\ell-1}^{(1)}+\eta_2c_{\ell-1}^{(2)}}{2\theta},
 \qquad c_0=1,
 \label{eq:smoothing}
\end{equation}
where $c_{\ell-1}^{(1)},c_{\ell-1}^{(2)}$ are independent copies of $c_{\ell-1}$, independent of the edge signs.

\begin{figure}[t]
 \centering
 \includegraphics[width=\linewidth]{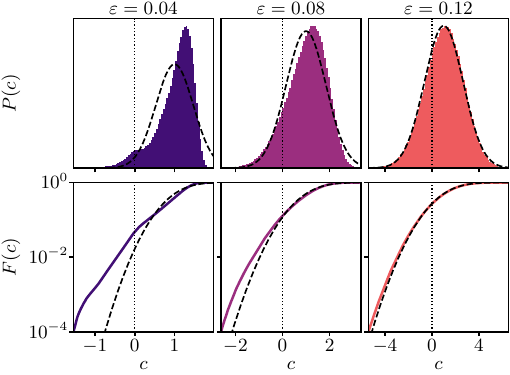}
 \caption{Law of the infinite-depth normalized census $c$, obtained by population dynamics on the recursion \eqref{eq:smoothing} ($2\times10^6$ samples, iterated until the effective $r^{-\ell}<10^{-4}$ and rescaled to unit empirical mean after each sweep). Top: density $P$; Bottom: cumulative distribution $F$, both compared with the matched Gaussian $\mathcal N(1,\gamma_\infty)$ (dashed). The dotted line marks the decision boundary, so that $F(0)$ is the census error. Left: at small noise the Gaussian underestimates the error and the left tail; Middle: at larger noise the distribution does not match but the census error is close to its Gaussian equivalent nonetheless as the cumulative distributions are similar at 0; Right: on approach to $\varepsilon_c\approx0.146$ the two laws merge.}
 \label{fig:census-popdyn}
\end{figure}

For $r>1$ ($\varepsilon < \varepsilon_c$), $c_\ell$ converges as $\ell\to\infty$ to a limit $c$, with $\E{c}=1$ and $\Var(c)=\gamma_\infty$~\cite{kesten1966limit,biggins1977martingale}, whose law is the distributional fixed point of Eq.~\eqref{eq:smoothing}. In practice, we sample it by population dynamics. A large pool of samples is updated by replacing each member with the right-hand side of Eq.~\eqref{eq:smoothing}, evaluated on two members drawn at random from the current pool and two fresh edge signs. Starting from $c_0=1$, after $\ell$ synchronous sweeps the pool approximates the law of $c_\ell$, up to finite-population error. The condition $r^{-\ell}\ll1$ controls the relative variance correction; sampling error must be checked separately. We show the resulting limiting law for different values of $\varepsilon$ in Fig.~\ref{fig:census-popdyn}.

The moments of $c$ follow order by order from this fixed point (Appendix~\ref{app:moments}). In particular, computing the third cumulant $\kappa_3$ gives the skewness
\begin{equation}
 \frac{\kappa_3}{\gamma_\infty^{3/2}} =-\frac{2(2\theta^2-1)^{3/2}}{(4\theta^2-1)(1-\theta^2)^{1/2}}<0.
\label{eq:skewness}
\end{equation}
The skewness is strictly negative for every $1<r<2$ and only vanishes as $r\to1$, on approach to KS. As such, even asymptotically, the limiting census is non-Gaussian, with excess weight on its left side (Fig.~\ref{fig:census-popdyn}).\footnote{The finite-depth census $c_\ell$ is likewise non-Gaussian; its exact law is given in Appendix~\ref{app:census-law}.} This excess weight stems from events at the top of the tree, such as a flip of one of the two root edges, as discussed below.

In the context of root classification, the deviation from Gaussianity matters most at the decision boundary. This is illustrated in the second row of Fig.~\ref{fig:census-popdyn}, which shows that although the distribution remains non-Gaussian, the cumulative distribution at the decision boundary may already be close to its Gaussian counterpart. At small $\varepsilon$, the impact of non-Gaussianity on the census prediction can be made precise. If the census were Gaussian with the same mean and variance, the root prediction error of the census estimator would be $\Phi(-1/\sqrt{\gamma_\infty})\le \e^{-1/(2\gamma_\infty)}$, with $\Phi$ the standard normal cumulative distribution function. Since $\gamma_\infty=4\varepsilon+O(\varepsilon^2)$, this is of order $\e^{-1/(8\varepsilon)}$ at small noise. The actual census error $\mathcal E^{(\ell)}_{\rm cens}=\Pp(c_\ell<0)+\tfrac{1}{2}\Pp(c_\ell=0)$ is instead of order $\varepsilon$. Indeed, if exactly one of the two root edges flips, which happens with probability $2\varepsilon(1-\varepsilon)$, the two halves of the tree carry opposite signals and the census is a fair coin. Conditioning on the two root edges, this gives (Appendix~\ref{app:census-law})
\begin{equation}
 \mathcal E^{(\ell)}_{\rm cens}=\varepsilon+(1-2\varepsilon)\,a_{\ell-1}
 \ge\varepsilon,
 \label{eq:census-floor}
\end{equation}
where $a_{\ell-1}$ is the error of the sum of two independent root-aligned depth-$(\ell-1)$ censuses, including half the tie probability. At small noise, no Gaussian description of the census can therefore reproduce its error, which it underestimates exponentially. For this reason, the learning theory we will develop for this data model will necessarily need to take into account this non-Gaussianity so as not to severely underestimate the error at small $\varepsilon$.

\subsubsection{Finite-depth effects}
\label{sec:census_finite_depth}

A peculiar feature of the data model at hand, and generically of tree-based data, is the strong separation between the sequence length $N$, which quickly becomes large, and the depth of the tree $\ell \sim \log N$ that is comparably much smaller. A correction algebraic in $\ell$ becomes a power of $1/\log N$, whereas a correction exponential in $\ell$ becomes a power of $1/N$. The census variance exhibits the latter behavior, with increasingly slow convergence near the reconstruction threshold. What is further atypical is that these corrections may be more or less important depending on the value of $\varepsilon$.

The most evident illustration comes from the variance of the normalized census, Eq.~\eqref{eq:gamma-inf}. There, the relative finite-depth correction is
\begin{equation}
 \frac{\gamma_\infty-\gamma_\ell}{\gamma_\infty}=r^{-\ell}=N^{-\log_2 r},
\end{equation}
a power law in $N$ with an exponent $\log_2r\in(0,1)$ that vanishes at KS. Exactly at KS, $\gamma_\ell=\ell/2=\tfrac{1}{2}\log_2N$ never converges, and the census SNR decays as $\sqrt{2/\log_2N}$. Expanding close to KS, $r-1\simeq4\sqrt2\,(\varepsilon_c-\varepsilon)$, so the asymptotic regime $\ell(r-1)\gg1$ is only reached outside the critical window
\begin{equation}
 |\varepsilon_c-\varepsilon|\lesssim\frac{1}{4\sqrt2\,\log_2N}.
\end{equation}
Note that this window is only symmetric at leading order. Beyond, the census behaves differently on the two sides of the transition, $\gamma_\ell$ converging for $r>1$ but growing as $r^{-\ell}$ for $r<1$. This explains why we later observe finite-size effects of different sizes on the two sides of the transition.

From this observation, we conclude that our learning theory will need to control some of the finite-size effects in order to adequately describe experiments performed even at large $N$.

\section{\texorpdfstring{A non-Gaussian learning theory}{A non-Gaussian learning theory}}
\label{sec:theory}

We now move on to a trained linear classifier---i.e., a perceptron---aiming at predicting the root. The weights $\bm w \in \mathbb{R}^N$ are obtained by empirical risk minimization (ERM) on a training set of $P$ independent leaves-root pairs $\{(\bm x^\nu,y^\nu)\}_{\nu=1}^P$,
\begin{equation}
 \hat{\bm w}=\argmin_{\bm w}\left\{\sum_{\nu=1}^P V\!\left(\frac{y^\nu\,\bm w\cdot\bm x^\nu}{\sqrt N}\right)+\frac{\lambda}{2}\|\bm w\|^2\right\},
 \label{eq:erm}
\end{equation}
with a convex loss $V$ and a ridge regularization of strength $\lambda>0$, our main example being the logistic loss $V(z)=\log(1+\e^{-z})$. The root of a fresh tree is then predicted as $\hat x_0=\sign(\hat{\bm w}\cdot\bm x)$, and we measure the test error $\mathcal E=\Pp(\hat x_0\neq y)$.

In the limit of infinite data at fixed dimension, the regularization in Eq.~\eqref{eq:erm} becomes negligible, since the loss is summed over samples, and $\hat{\bm w}$ therefore approaches the minimizer of the unregularized population loss. This minimizer is in fact the census, which is thus the natural reference point. Indeed, exchanging the two subtrees hanging from any node leaves the joint law of $(\bm x,y)$ invariant, and composing such exchanges maps any leaf onto any other. Since we take a convex $V$, averaging a minimizer of the population loss over these exchanges yields a minimizer with equal weights on all leaves. Thus the census is the population logistic predictor, and more generally the minimizer of the population loss for any convex loss decreasing at zero.\footnote{For $0<\varepsilon<1/2$, the leaves are positively correlated with the root, so that for $V$ decreasing at zero the population loss also decreases from $\bm w=0$ along the census direction, ensuring that the null vector is not selected.} We now wish to understand what occurs when considering finitely many samples, where we naturally expect fluctuations of $\hat{\bm w}$ around the census direction $\ee$.

\subsection{Asymptotic results in the proportional regime}
\label{sec:proportional}

As is standard in the statistical mechanics analysis of empirical risk minimization, we start by considering the asymptotic proportional regime, with $N,P\to \infty$ at fixed $\alpha=P/N$. The typical properties of the minimizer \eqref{eq:erm} are obtained by studying the partition function
\begin{equation}
 Z_\beta=\int\dd\bm w\,\exp\left(-\beta\Big[\sum_{\nu=1}^P V(z^\nu)+\frac{\lambda}{2}\|\bm w\|^2\Big]\right),
\end{equation}
where $z^\nu=y^\nu\bm w\cdot\bm x^\nu/\sqrt N$ are the training margins, and the associated free energy
\begin{equation}
 f_N=-\lim_{\beta\to\infty}\frac{1}{\beta N}\langle \log Z_\beta \rangle,
 \label{eq:free-energy}
\end{equation}
where the zero-temperature limit $\beta\to\infty$ selects the minimizer and the angular bracket refers to the average over the training samples. We compute the average over training sets with the replica method \cite{mezard1987spin,mezard2009information}, which gives $f_N$ at leading order as $N,P\to\infty$ at fixed $\alpha$, in the replica-symmetric ansatz appropriate for convex losses, and relegate the derivation to Appendix~\ref{app:replica}.

In the standard approach, the average over the data is straightforward, since the inputs are assumed to be Gaussian, possibly with a structured covariance. Here however, we have emphasized that the relevant projection of the data is not Gaussian in general. In order to account for this non-Gaussianity, the crucial step here is to decompose the weights along and orthogonal to the census direction, $\bm w=\sqrt N m\,\ee+\bm w_\perp$ with $\ee\cdot\bm w_\perp=0$, where $m=\1\cdot\bm w/N$ is the projection of the weights along the census direction.
As the signal along $\ee$ grows with the depth, $y\ee\cdot\bm x=s_\ell c_\ell$ with $s_\ell^2=r^\ell$, we introduce the rescaled signal overlap $\mb=s_\ell m$, which remains of order one as $\ell\to\infty$. The margin of a sample then splits exactly as
\begin{equation}
 z=\mb\,c_\ell+\xi,\qquad \xi=\frac{y \bm w_\perp\cdot\bm x}{\sqrt N}.
 \label{eq:margin-split}
\end{equation}
This isolates the non-Gaussianity of the direction carrying the label in the scalar $c_\ell$, with the exact law of Sec.~\ref{sec:census}. The transverse field $\xi$ has zero mean and variance $\qp=\bm w_\perp^\top\bm\Sigma_\perp\bm w_\perp/N$, with $\bm\Sigma_\perp=(\mathbf I-\ee\ee^\top)\bm\Sigma(\mathbf I-\ee\ee^\top)$ the transverse covariance, whose spectrum is $\rho_{\perp,\ell}$, and is uncorrelated with $c_\ell$ because $\ee$ is an eigenvector of $\bm\Sigma$. We approximate it as a Gaussian variable independent of $c_\ell$. This is the central approximation of the theory.

The replica-symmetric saddle point reduces the free energy computation to an optimization problem with few order parameters. In addition to the rescaled overlap $\mb$, these are the overlap $\qp$ and the susceptibility $\chi$ of the transverse weights---defined from the overlaps $Q^\perp_{ab}=\bm w^{a\top}_\perp\bm\Sigma_\perp\bm w^b_\perp/N$ of two replicas as $Q^\perp_{a\neq b}=\qp$ and $\chi=\lim_{\beta\to\infty}\beta(Q^\perp_{aa}-\qp)$---together with their conjugates $\hat q$ and $\hat\chi$. The resulting saddle-point value $f^{(\ell)}$ takes the standard form
\begin{equation}
 f^{(\ell)}=\extr_{\mb,\qp,\chi,\hat q,\hat\chi}\big[g_I+g_S+\alpha\,g_E\big],
 \label{eq:action}
\end{equation}
with the interaction, entropic and energetic terms
\begin{equation}
\begin{aligned}
 g_I&=\frac{\hat q\chi-\hat\chi\qp}{2},\quad
 g_S=\frac{\lt_\ell\mb^2}{2}-\frac{\hat q}{2}\Big\langle\frac{\mu}{\lambda+\mu\hat\chi}\Big\rangle_{\rho_{\perp,\ell}},\\
 g_E&=\Big\langle\min_z\Big[\frac{(z-h)^2}{2\chi}+V(z)\Big]\Big\rangle_{c,\zeta}.
\end{aligned}
\label{eq:g-terms}
\end{equation}
Here $\langle\cdot\rangle_{\rho_{\perp,\ell}}$ denotes the average over the eigenvalues $\mu$ of $\bm\Sigma_\perp$, with density $\rho_{\perp,\ell}$, and $\langle\cdot\rangle_{c,\zeta}$ the average over the census law and an independent standard Gaussian $\zeta$, with $h=\mb c_\ell+\sqrt{\qp}\,\zeta$ the field of a sample. Since the signal along $\ee$ grows as $s_\ell$, the ridge acts on the census coordinate with the rescaled strength $\lt_\ell=\lambda/s_\ell^2=\lambda r^{-\ell}$, whereas it keeps its bare value $\lambda$ on the transverse directions. Most transverse eigenvalues are of order one, although the largest ones grow with depth for $r>1$. We retain the explicit dependence on the tree depth $\ell$ through the rescaled ridge $\lt_\ell$, the law of $c_\ell$ and the spectrum $\rho_{\perp,\ell}$, before taking their infinite-depth limits below. Writing $z_\star(h)=\argmin_z[(z-h)^2/(2\chi)+V(z)]$ for the minimizer in $g_E$, which is unique for convex $V$, stationarity gives
\begin{equation}
\begin{aligned}
 \chi&=\Big\langle\frac{\mu}{\lambda+\mu\hat\chi}\Big\rangle_{\rho_{\perp,\ell}},\\
 \hat q &=\alpha\big\langle V'(z_\star)^2\big\rangle_{c,\zeta},\\
 \qp&=\hat q\,\Big\langle\frac{\mu^2}{(\lambda+\mu\hat\chi)^2}\Big\rangle_{\rho_{\perp,\ell}},\\
 \hat\chi&=\alpha\Big\langle\frac{V''(z_\star)}{1+\chi V''(z_\star)}\Big\rangle_{c,\zeta},\\
 \lt_\ell\,\mb&=-\alpha\big\langle c_\ell V'(z_\star)\big\rangle_{c,\zeta}.
\end{aligned}
\label{eq:saddle}
\end{equation}
Assuming that the margin of a fresh sample is also distributed as $h$ with transverse Gaussian fluctuations, for $\qp>0$ the predicted test error follows from the saddle-point equations as
\begin{equation}
 \mathcal E=\left\langle\Phi\left(-\frac{\mb c_\ell}{\sqrt{\qp}}\right)\right\rangle_{c}.
 \label{eq:test-error}
\end{equation}
A moment-matched Gaussian census would amount to taking $c_\ell\sim\mathcal N(1,\gamma_\ell)$, in which case Eqs.~\eqref{eq:saddle} reduce to those of a Gaussian mixture with the same covariance~\cite{loureiro2021learning} (Appendix~\ref{app:replica}). In the following, we refer to the saddle point of Eqs.~\eqref{eq:saddle}, which combines the exact law of the census with Gaussian transverse averages, as a \textit{hybrid} theory.

\begin{figure}
 \centering
 \includegraphics[width=\linewidth]{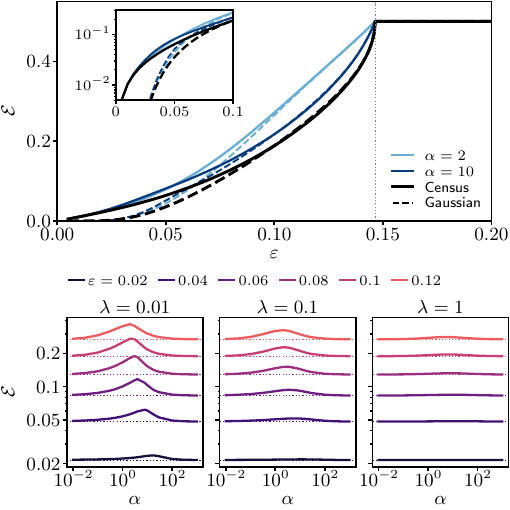}
 \caption{Test error predicted by the infinite-depth limit of the replica theory. Top: error as a function of $\varepsilon$ at $\lambda=0.01$, for $\alpha = 2,10$, from the non-Gaussian saddle (solid) and from the Gaussian saddle (dashed), compared with the census error (black, solid) and its Gaussian counterpart $\Phi(-1/\sqrt{\gamma_\infty})$ (black, dashed). The dotted line marks $\varepsilon_c$, beyond which the leaves carry no information on the root at infinite depth. The inset shows the same curves for $\varepsilon\le0.1$ on a logarithmic scale. Bottom: error as a function of $\alpha$ for three values of the ridge $\lambda$ and several values of $\varepsilon$, with the corresponding census errors dotted.}
 \label{fig:saddle-inf}
\end{figure}

For $1<r<2$ and fixed $\alpha$, the three depth-dependent inputs converge, $c_\ell\to c$, $\rho_{\perp,\ell}\to\rho$ and $\lt_\ell\to0$, which defines the infinite-depth free energy $f=\lim_{\ell\to\infty}f^{(\ell)}$, in which the last of Eqs.~\eqref{eq:saddle} becomes $\langle c\,V'(z_\star)\rangle_{c,\zeta}=0$. In practice, we sample the fixed-point law of $c$ by population dynamics, as in Fig.~\ref{fig:census-popdyn}, and evaluate the averages over $c$ on the resulting population. The spectral averages only involve bounded resolvents, and thus converge even though the moments of $\rho$ may diverge, as anticipated in Sec.~\ref{sec:correls}. We truncate them at a level $K$, with an error that is exponentially small in $K$ (Appendix~\ref{app:truncation}). As the full and transverse spectra only differ by the weight of their top eigenvalue $\mu_\ell$, this truncated spectrum is common to both, and to their limit $\rho$.

The resulting asymptotic test error for different values of $\alpha, \varepsilon, \lambda$ is shown in Fig.~\ref{fig:saddle-inf}. There, we also show the equivalent Gaussian saddle, obtained by replacing the census law by $\mathcal N(1,\gamma_\infty)$. As expected from the discussion on the non-Gaussianity of the census above, the Gaussian approximation agrees with the non-Gaussian one close to KS, but underestimates the error by orders of magnitude at small noise. The non-Gaussian saddle instead remains close to the census error at both ends, $\alpha\to0$ and $\alpha\to\infty$. At small ridge, the error is non-monotonic in between, with a maximum at $\alpha$ of order one to ten that moves to larger $\alpha$ as $\varepsilon$ decreases. This maximum, which is associated with the trained weights fitting transverse noise directions, resembles sample-wise double descent~\cite{nakkiran2019more} and can be smoothed out by increasing the regularization~\cite{nakkiran2021optimal}. We do not identify a sharp interpolation threshold at finite $\lambda$.

The fact that the error approaches the census benchmark as $\alpha\to0$ is surprising at first and points towards a sub-proportional regime. Indeed, since the limit $\ell\to\infty$ is taken first, $\alpha\to0$ still corresponds to a diverging number of samples $P=\alpha N$, which suggests that far fewer than $N$ samples already bring the error close to the census error. Before exploring that regime, we investigate the effect of keeping a finite $\ell$ in the saddle.

\subsection{Finite-depth trees}
\label{sec:finite-size}

So far, we have studied the infinite-depth asymptotics. As described in Sec.~\ref{sec:census_finite_depth}, however, we expect that these will poorly describe finite-$N$ experiments due to the strong finite-depth effects inherited from the hierarchy.

Three effects must be distinguished: the depth dependence of the hybrid model, corrections to its large-$N$ replica saddle, and the discrepancy between the hybrid model and the broadcast data from the Gaussian assumption on transverse fluctuations. We quantify the first of these. The calculation below does not establish the precise $O(1/N)$ remainder for the replica approximation or a rate for the Gaussian approximation of the transverse field.

We start with the covariance spectrum. We first notice that the integrands of the spectral averages in Eq.~\eqref{eq:saddle} increase with $\mu$ but saturate,
\begin{equation}
 0\le\frac{\mu}{\lambda+\mu\hat\chi}\le\frac{1}{\hat\chi},\qquad
 0\le\frac{\mu^2}{(\lambda+\mu\hat\chi)^2}\le\frac{1}{\hat\chi^2},
 \label{eq:resolvent-saturation}
\end{equation}
such that the contribution of a large eigenvalue is set by its weight in the spectrum rather than by its magnitude. Moreover, $\rho_{\perp,\ell}$ is exactly the limiting density $\rho$ truncated at height $\ell$, so the two naturally share the eigenvalues $\mu_k$ with weights $2^{-k}$ for $k\le\ell$, and only differ by the mass $2^{-\ell}$ that $\rho$ places on the larger eigenvalues $\mu_k$, $k>\ell$. Together with the saturation, the finite-depth spectral averages therefore differ from their limits by at most $2^{-\ell}\hat\chi^{-j}=O(1/N)$, $j=1,2$ (Appendix~\ref{app:truncation}). Thus, at fixed positive $\hat\chi$, replacing the finite-depth spectrum by the limiting one introduces an $O(1/N)$ error in these spectral averages, of the same order as the corrections expected from evaluating the free energy at the saddle point. %
Note that this no longer holds as $\alpha\to0$, since $\hat\chi\propto\alpha$ [Eq.~\eqref{eq:saddle}], as we shall see in Sec.~\ref{sec:sample_complexity}.

We now turn to the census averages. Since an infinite tree is a depth-$\ell$ tree whose leaves each root an independent infinite subtree, we can write $c=c_\ell+\delta_\ell$ on the same tree, where $\delta_\ell$ collects the fluctuations of these subtrees, with $\E{\delta_\ell\mid c_\ell}=0$ and $\E{\delta_\ell^2}=\gamma_\infty-\gamma_\ell=\gamma_\infty r^{-\ell}$, so that the first-order term cancels in any smooth average and expanding to second order gives
\begin{equation}
 \big|\langle g(c)\rangle_c-\langle g(c_\ell)\rangle_c\big|\le\tfrac{1}{2}\|g''\|_\infty\,\gamma_\infty\,r^{-\ell},
 \label{eq:census-depth-error}
\end{equation}
for any twice differentiable $g$ with bounded second derivative. For the logistic loss, this applies to the census averages of Eqs.~\eqref{eq:saddle} and to the test error, Eq.~\eqref{eq:test-error}. %
The census therefore produces corrections of order $r^{-\ell}=N^{-\log_2r}$ in these averages, as does the rescaled ridge $\lt_\ell=\lambda r^{-\ell}$. These estimates give, for $1<r<2$ and fixed $\alpha>0$,
\begin{equation}
 f^{(\ell)}=f+O(r^{-\ell}),
 \label{eq:fN-expansion}
\end{equation}
and the same order for the predicted test error. These depth corrections are the essential ones here, since they decay more slowly than the $O(1/N)$ spectral and saddle point corrections as $r<2$. They are captured by solving Eqs.~\eqref{eq:saddle} with the finite-depth inputs $c_\ell$, $\rho_{\perp,\ell}$ and $\lt_\ell$, which defines the finite-depth saddle used in the following.

We compare the finite-depth saddle with empirical risk minimization on broadcast trees in Fig.~\ref{fig:finite-depth}; numerical implementation details are given in Appendix~\ref{app:numerics}. The agreement is very good at all depths, including above the threshold, where the error approaches $1/2$ only slowly with $\ell$. Note that the agreement also holds at the level of the order parameters of the trained classifiers, see Appendix~\ref{app:orderparams}. The finite-depth hybrid saddle remains well-defined on either side of the threshold. The restriction $r>1$ is needed for the normalized census to have the finite-variance infinite-depth limit used in the preceding convergence analysis. The infinite-depth saddle, instead, fails to describe trees of accessible depth close to $\varepsilon_c$, as we anticipated in Sec.~\ref{sec:census_finite_depth}. The bottom row of Fig.~\ref{fig:finite-depth} shows the corresponding learning curves as a function of $\alpha$. At small depth, the error now decreases monotonically with $\alpha$, as in the standard picture of learning. At the moderate depths we considered, the minimum error for $\varepsilon$ close to $\varepsilon_c$ is reached as $\alpha \to \infty$ rather than equally as $\alpha \to 0$, there again as in traditional learning settings.

\begin{figure}
 \centering
 \includegraphics[width=\linewidth]{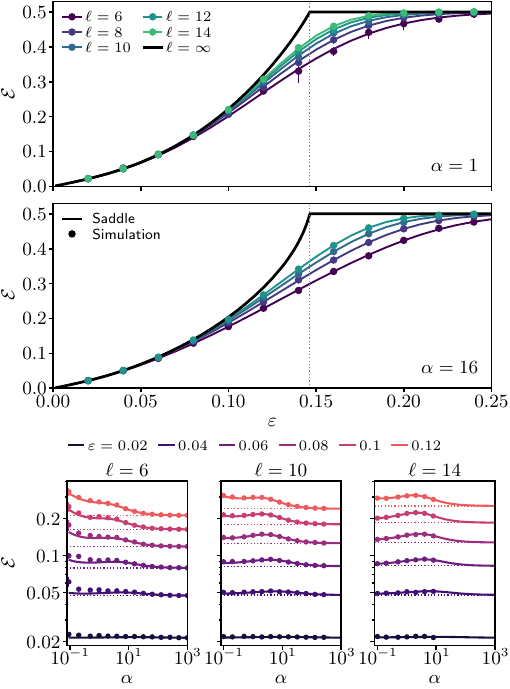}
 \caption{Test error of logistic regression on finite-depth trees, $\lambda=0.1$. Top two panels: error as a function of $\varepsilon$ for $\alpha=1$ and $\alpha=16$ at depths $\ell=6$--$14$, from the finite-depth hybrid saddle (lines) and from empirical risk minimization on broadcast trees (dots, approximate $95\%$ intervals, see Appendix~\ref{app:numerics}), compared with the infinite-depth saddle (black). The dotted line marks $\varepsilon_c$. Bottom: error as a function of $\alpha$ at depths $\ell=6$, $10$ and $14$ for several values of $\varepsilon$ (same colors as Fig.~\ref{fig:saddle-inf}), from the finite-depth saddle (lines) and from simulations (dots), with corresponding finite-depth census errors dotted.}
 \label{fig:finite-depth}
\end{figure}

\subsection{Sub-proportional regime}
\label{sec:sample_complexity}

So far, we observed that at infinite depth the error of the empirical risk minimization matches the census error already as $\alpha\to0$. This observation points towards the census direction being learned from a vanishing fraction of samples, $P\ll N$. Even at finite $\ell$, it appears that the census accuracy is quickly reached for small $\alpha$. We now precisely identify the scale at which this happens and the effect of the finite tree depth in this context.

To identify the scale of learning, we first consider the signed mean of the training set,
\begin{equation}
\hat{\bm v}=\frac{1}{P} \sum_\nu y^\nu\bm x^\nu.
\label{eq:signed_mean}
\end{equation}
Each training sample contributes to $\hat{\bm v}$ through its signed leaves $y^\nu\bm x^\nu$. Along the signal direction, this contribution is the signed census of Sec.~\ref{sec:census}, $y^\nu\ee\cdot\bm x^\nu=s_\ell c_\ell^\nu$, of mean $s_\ell$. These contributions add up coherently over the training set, and writing $\bar c_P=P^{-1}\sum_\nu c_\ell^\nu$ the empirical average of the census---of mean one and variance $\gamma_\ell/P$---the rescaled overlap of $\hat{\bm v}$ is exactly $\mb=s_\ell^2\bar c_P/\sqrt N$. For $P\gg\gamma_\ell$ the average concentrates, $\bar c_P\simeq1$, so that the overlap is deterministic, $\mb\simeq s_\ell^2/\sqrt N$. In the transverse directions, on the other hand, the contributions have zero mean and covariance $\bm\Sigma_\perp$, so that their average $\hat{\bm v}_\perp$ has covariance $\bm\Sigma_\perp/P$. Thus $\E{\qp}=\Tr\bm\Sigma_\perp^2/(NP)=B_\ell/P$, where the expectation is over the training set and
\begin{equation}
 B_\ell=\frac{1}{N}\Tr\bm\Sigma_\perp^2=\sum_{k=1}^{\ell-1}2^{-k}\mu_k^2+2^{-\ell}\mu_\ell^2
 \label{eq:Bell}
\end{equation}
is the second moment of the transverse spectrum. If $\bar c_P$ and $\qp$ concentrate about these values, and the transverse field is Gaussian and independent of the census as in Sec.~\ref{sec:proportional}, the test error of Eq.~\eqref{eq:test-error} depends on the ratio $\mb^2/\qp\simeq Ps_\ell^4/(NB_\ell)$. For $1<r<2$, this identifies the onset of an order-one advantage over chance ($1/2-\mathcal E$) when the ratio is of order one, which corresponds to a number of samples of order
\begin{equation}
 P_\star=\frac{NB_\ell}{s_\ell^4}.
 \label{eq:Pstar}
\end{equation}
The behavior of $B_\ell$ follows from the spectral analysis of Sec.~\ref{sec:correls}. The terms of Eq.~\eqref{eq:Bell} importantly scale as $(r^2/2)^k$, so that $B_\ell$ remains finite for $r<\sqrt2$, with the limit $B_\infty=(4-r^2)/[2(2-r^2)]$. Precisely at $r=\sqrt2$, $B_\ell\asymp\ell$, meaning that each sufficiently large spectral level contributes an amount of order one, so the divergence accumulates over the hierarchy. Finally, for $\sqrt2<r<2$, $B_\ell\asymp r^{2\ell}/N$ is dominated by the top eigenvalues. Bringing these results together with the expressions $s_\ell^2=r^\ell$ and $N=2^\ell$, the sample scale therefore behaves as
\begin{equation}
 P_\star\asymp
 \begin{cases}
  N^{\omega(\varepsilon)}, & 1<r<\sqrt2,\\
  \log N, & r=\sqrt2,\\
  1, & \sqrt2<r<2,
 \end{cases}
 \qquad \omega(\varepsilon)=1-2\log_2r,
 \label{eq:sample-exponent}
\end{equation}
where $r=\sqrt2$ corresponds to $\varepsilon_\star=(1-2^{-1/4})/2\approx0.080$. In plain words, the exponent tends to $\omega=1$ on approach to KS from below and to $\omega=0$ on approach to $\varepsilon_\star$, where the marginal scale is $\log N$. Below $\varepsilon_\star$, $P_\star$ remains of order one. At KS itself the census error tends to $1/2$, so the formal scale $P_\star\asymp N$ does not imply nontrivial reconstruction. The sample scale is thus continuously tuned by the strength of the hierarchical correlations. Note that this scale is not a fundamental limit, since the census itself requires no training. It is rather the number of samples needed to find the census direction by the estimators considered here, which have no prior preference for it. As shown below, it does not depend on the specific loss or regularization taken.

Interestingly, $\varepsilon_\star$ can itself be read as a KS threshold. This follows from the fact that the product of two independent broadcast trees, taken node by node with each tree aligned with its own root, is itself a broadcast tree. It equals $+1$ at the root and flips along an edge when exactly one of the two trees does, i.e.\ with probability $\varepsilon'=2\varepsilon(1-\varepsilon)$, so that $\theta'=\theta^2$ and $r'=r^2/2$. Such products are what the signed mean relies on to classify a test tree. Its margin, Eq.~\eqref{eq:margin-split} with $\bm w=\hat{\bm v}$, is indeed proportional to $\sum_\nu\sum_iu^\nu_i$, where $u^\nu_i=(yx_i)(y^\nu x^\nu_i)$ are the leaves of the \textit{overlap tree} formed by the test tree and the training tree $\nu$, so that it is the sum of the censuses of $P$ overlap trees. The pair correlations of an overlap tree are the squares of those of the original tree, $\E{u_iu_j}=\E{x_ix_j\mid y}^2$, so that the variance of its normalized census, defined as for the original tree, is $\gamma'_\ell=P_\star+\gamma_\ell(2+\gamma_\ell)$, that is $P_\star$ up to a term of order one. The sample scale $P_\star$ is thus the number of overlap trees that must be averaged to resolve the unit mean of their census, and the three regimes of Eq.~\eqref{eq:sample-exponent} are those of the overlap tree on either side of its own KS threshold $r'=1$, i.e.\ $r=\sqrt2$. For $\varepsilon>\varepsilon_\star$, a single training example therefore carries vanishing information as $\ell\to\infty$, whereas for $\varepsilon<\varepsilon_\star$ it classifies better than chance at any depth, with an error equal to the census error of the overlap tree.

For $\varepsilon_\star<\varepsilon<\varepsilon_c$, the learning curve on this scale can be obtained exactly for the signed mean of Eq.~\eqref{eq:signed_mean}. Taking $P/P_\star\to\tau\in(0,\infty)$ as $\ell\to\infty$, plugging in the values of $\mb$ and $\qp$ derived above in Eq.~\eqref{eq:test-error} gives
\begin{equation}
 \mathcal{E}_{\rm on}(\tau)=\big\langle\Phi(-\sqrt\tau\,c)\big\rangle_c.
 \label{eq:onset}
\end{equation}
Note that in this regime, the Gaussian census result $\Phi(-\sqrt{\tau/(1+\gamma_\infty\tau)})$ is numerically very close to this prediction, as non-Gaussian effects become noticeable only at small $\varepsilon$.

For $\varepsilon_c<\varepsilon<1/2$, the scale in Eq.~\eqref{eq:Pstar} instead gives $\tau=P/P_\star\propto\alpha r^{2\ell}\to0$ at fixed $\alpha$. In the signed-mean approximation, the finite-depth expression $\langle\Phi(-\sqrt\tau c_\ell)\rangle$ then predicts an advantage of order $\sqrt\tau\propto\sqrt\alpha\,r^\ell$. This vanishes slowly near KS, consistently with the visible finite-depth advantage in Fig.~\ref{fig:finite-depth}. Equation~\eqref{eq:onset}, which uses a limiting unit-mean census of finite variance, does not apply for $r\le1$.

\begin{figure}
 \centering
\includegraphics[width=\linewidth]{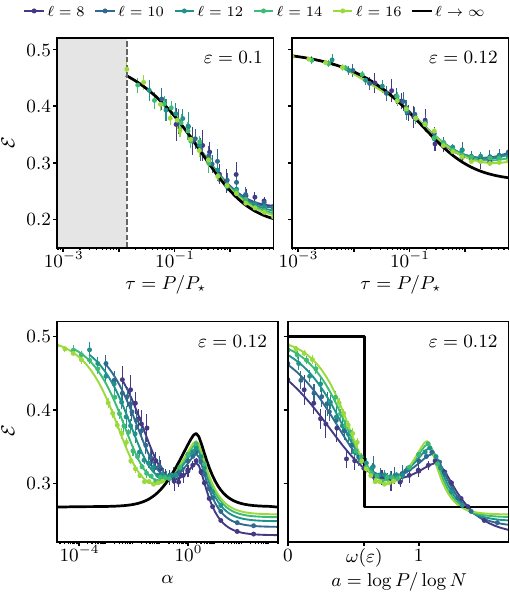}
 \caption{Sub-proportional regime at $\lambda=0.01$, for depths $\ell=8$--$16$ from simulations on broadcast trees (dots, approximate $95\%$ intervals, see Appendix~\ref{app:numerics}) and the finite-depth saddle (lines). Top: test error as a function of $\tau=P/P_\star$ at $\varepsilon=0.1$ (left) and $\varepsilon=0.12$ (right), compared with the onset limit $\mathcal{E}_{\rm on}(\tau)$ (black). On the left, the shaded region lies below the smallest value of $\tau$ reachable at $\ell=16$, with a single training sample. Bottom: test error at $\varepsilon=0.12$ as a function of $\alpha=P/N$ (left), with the infinite-depth saddle in black, and of $a=\log P/\log N$ (right), with the infinite-depth step at $a=\omega(\varepsilon)$ in black, from $1/2$ to the limiting census error. The peak of the proportional regime shrinks to the single point $a=1$ in this limit and is not drawn.}
 \label{fig:subext}
\end{figure}

Importantly, the finite-depth saddle of Sec.~\ref{sec:finite-size} makes relevant predictions for the empirical risk minimization in this sub-proportional regime. For $1<r<\sqrt2$, fixed $\lambda>0$ and $P=\tau P_\star$ with fixed $\tau>0$, the sample ratio $\alpha$ vanishes. To make contact with this regime, we therefore solve the saddle-point equations~\eqref{eq:saddle} at leading order in $\alpha$, keeping each order parameter at its first non-vanishing order. The conjugates $\hat q$ and $\hat\chi$ are of order $\alpha$. The limit can be taken in the spectral averages by dominated convergence, since the first two moments of $\rho$ are finite, giving $\chi\simeq\langle\mu\rangle_{\rho_{\perp,\ell}}/\lambda=1/\lambda$, of order one, and $\qp\simeq\hat qB_\ell/\lambda^2$, of order $\alpha$. The cavity fields $h=\mb c_\ell+\sqrt{\qp}\,\zeta$ are small provided $\mb$ is small as well. At leading order, the minimizer $z_\star$ in $g_E$, which satisfies $z_\star=h-\chi V'(z_\star)$, is thus the same margin $z_0$ for all samples, solution of $\lambda z_0=-V'(z_0)$. The averages over $c_\ell$ and $\zeta$ in the equations for $\hat q$ and $\mb$ can then be performed at $z_\star=z_0$, which gives
\begin{equation}
 \hat q\simeq\alpha V'(z_0)^2,\quad
 \lt_\ell\,\mb\simeq-\alpha V'(z_0)\langle c_\ell\rangle=\alpha|V'(z_0)|,
 \label{eq:saddle-onset}
\end{equation}
using $\langle c_\ell\rangle=1$ and the decreasing nature of $V$. The equation for $\mb$ is where keeping the finite depth is essential, as both of its sides are small. The right-hand side, which stems from the loss in $\alpha g_E$, is of order $\alpha$, while the left-hand side, which stems from the regularization in $g_S$, carries the vanishing rescaled ridge $\lt_\ell=\lambda r^{-\ell}$. Their balance gives $\mb\simeq\alpha r^\ell|V'(z_0)|/\lambda$, which together with $\qp\simeq\alpha V'(z_0)^2B_\ell/\lambda^2$, yields
\begin{equation}
 \frac{\mb^2}{\qp}\simeq\frac{\alpha r^{2\ell}}{B_\ell}=\frac{P}{P_\star}=\tau,
\end{equation}
in which the loss and the ridge drop out, as they only enter through the common factor $|V'(z_0)|/\lambda$. This also verifies $\mb=O(r^{-\ell})\to0$ on the onset scale. The saddle therefore predicts the same leading test error as the signed mean, $\mathcal E_{\rm on}(\tau)$. Again, we emphasize that this regime is only accessible because the finite-depth inputs are retained: setting $\lt_\ell=0$ at fixed $\alpha$, as in the infinite-depth limit of Fig.~\ref{fig:saddle-inf}, before sending $\alpha\to0$, discards the balance between signal and rescaled regularization that is responsible for the onset.

Figure~\ref{fig:subext} tests these predictions against numerical experiments performed in the data-scarce regime. When plotted against $\tau$ (top row), the simulations approach the onset curve at small $\tau$. At larger $\tau$, they eventually level off above it, as the empirical risk minimizer starts to fit the training set, ahead of the overfitting peak of the proportional regime. Moreover, the experiments show that the ERM is well described by the finite-depth saddle down to sample ratios $\alpha$ of order $10^{-4}$. As a function of the exponent $a=\log P/\log N$, the finite- $\ell$ error slowly approaches the sharp step at $a=\omega(\varepsilon)$ predicted as $\ell\to\infty$, where the census accuracy is reached, followed by the census plateau up to the proportional regime at $a=1$.

For $0<\varepsilon<\varepsilon_\star$, instead, $P_\star$ remains finite as $\ell$ grows. This identifies an order-one sample scale for resolving signal, rather than for consistently recovering the census direction or its accuracy. At fixed $P$, the training census and the dominant transverse modes retain fluctuations, so the preceding self-averaging onset formula is not justified. A description of this regime must retain training-set fluctuations, although their distribution and the mean error can likely still be studied. The marginal case $r=\sqrt2$ likewise requires a separate analysis of the logarithmic sample scale.

\begin{figure}[t]
 \centering
\includegraphics[width=\linewidth]{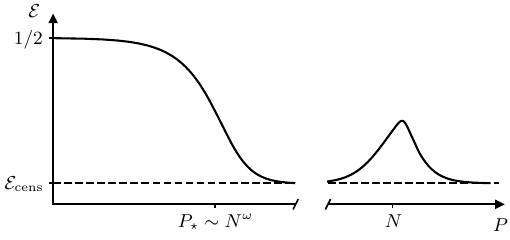}
 \caption{Learning curve at large depth for $\varepsilon_\star<\varepsilon<\varepsilon_c$, as a function of the number of samples $P$ on a logarithmic scale. Around $P_\star\sim N^\omega$, the error decreases from $1/2$ to the census error following the onset limit of Eq.~\eqref{eq:onset}; it then remains on the census plateau up to the proportional regime $P\sim N$, where the fitting of the transverse fluctuations produces a regularization-dependent peak, before the census error is recovered. Both parts are the $\ell\to\infty$ limits on their own scale, in $\tau=P/P_\star$ and $\alpha=P/N$ respectively, here at $\varepsilon=0.12$ and $\lambda=0.01$. The break in the axis hides the plateau, whose extent grows as $N^{1-\omega}$.}
 \label{fig:schematic}
\end{figure}

Altogether, for $\varepsilon_\star<\varepsilon<\varepsilon_c$ and a sufficiently small fixed ridge, the theory predicts the following learning curve at large depth, shown in Fig.~\ref{fig:schematic}. Around $P\sim P_\star$, the error decreases from $1/2$ to the census error, following $\mathcal E_{\rm on}$. In the proportional regime $P=\alpha N$, the fitting of the transverse fluctuations then produces the peak of Fig.~\ref{fig:saddle-inf}, at $\alpha$ of order one that is smoothed as $\lambda$ increases, before the census error is recovered as $\alpha\to\infty$. The learning curve is thus a double descent, but a scale-separated one, as the first descent is fixed by the statistics of the data alone and takes place at $P\sim N^\omega$, parametrically below the peak at $P\sim N$, which instead depends on the regularization, and in particular on whether $\lambda$ or $\lambda/\alpha$ is held fixed as $P$ grows. Between the two scales, the error sits on the census plateau, the $\alpha\to0$ limit of Fig.~\ref{fig:saddle-inf}. The separation between the two scales, however, only grows as $N^{1-\omega}$. At accessible depths, the first descent, the peak and the second descent are all visible in Fig.~\ref{fig:subext}, where they are accurately described by the finite-depth saddle, but between the two descents the error only reaches a minimum above the census error, rather than a plateau. Close to KS, $\omega\to1$ and the onset merges with the proportional regime, where the finite-depth effects of Sec.~\ref{sec:finite-size} take over.

\section{Discussion}
\label{sec:discussion}

In this work, we have studied an analytically tractable model of hierarchical data with the tools of statistical mechanics. The simplicity of the tree-based broadcast data model allowed us to develop and test a hybrid theory for the learning curves of a linear classifier from sub-proportional to proportional numbers of samples, accounting for the non-Gaussianity, the finite-size effects and the reduced sample complexity that stem directly from the hierarchy. We believe that this constitutes a first step towards statistical-mechanics descriptions of learning from inherently non-Gaussian, structured data, in the spirit of the program outlined by \citet{mezard2024spin}.

The fact that an accurate description at small $\varepsilon$ requires keeping the non-Gaussian law of the census in the saddle-point equations may appear surprising, given the many results establishing the robustness of Gaussian equivalence in high dimension~\cite{goldt2022gaussian,hu2023universality,montanari2022universality,loureiro2021curves,dandi2023universality}. Such Gaussian-equivalence results impose model-specific hypotheses, including control of projections along the relevant directions and suitable concentration properties. They do not imply universality for arbitrary correlated inputs. In the reconstructible broadcast model, $y\ee\cdot\bm x/s_\ell=c_\ell$ remains non-Gaussian even at infinite depth, and matching the input means and covariances does not reproduce its classification error. This is consistent with the mechanisms of universality breakdown discussed in Refs.~\cite{pesce2023gaussian,mai2025breakdown}. Our hybrid theory, which keeps the exact law along $\ee$ and treats the transverse field as an independent Gaussian, is analogous to the asymptotic descriptions of Refs.~\cite{mai2025breakdown,yaakoubi2026characterization} and to conditional Gaussian equivalence~\cite{cui2024asymptotics,wen2025when}. Our setting is not directly covered by these results, and we have not proven that the transverse field becomes Gaussian here. More generally, this points to a central tension for genuinely hierarchical data: a Gaussian description retains the hierarchical covariance, but washes out the non-Gaussian statistics that the hierarchy imprints on the few directions relevant to the task. The hybrid theory we propose suggests a way out, by isolating these directions and treating the rest as Gaussian, without losing solvability.

In this sense, the broadcast model, despite its simplicity and an unusually strong signal growing with the dimension, offers a first setting where this approach can be tested and pushed towards more complex architectures. The census, to which linear empirical risk minimization converges at large sample ratio, can have a larger error than the Bayes-optimal prediction of belief propagation, thereby leaving room to study how deeper architectures may approach Bayes optimality. Random features models, which learn the target degree by degree~\cite{aguirre2025random}, would be a natural first step. One possibility is that nonlinear corrections to the census require a higher power of $N$ samples, producing a further, separated descent towards the Bayes error. Establishing whether this occurs requires the spectral weights and target alignment of the chosen feature map; it does not follow from the linear analysis alone. Genuine feature learning, for instance through layerwise training~\cite{ren2026provable}, or convolutional architectures matched to the tree \cite{cagnetta2024how}, could instead bring this third descent to smaller numbers of samples. It should be noted, however, that the gain of belief propagation over the census is small in this model, meaning that such a third descent may be difficult to observe in numerical experiments, and would rather call for an asymptotic theory.

Another direction where having such a simple hierarchical data model may prove valuable is self-supervised learning and generative modeling. Indeed, it would be interesting to understand whether the strong signal carried by the root makes the generation of new sequences comparatively easy, or whether the weak, non-Gaussian correlations within the tree are instead hard to reproduce. Generative diffusion has notably been studied with hierarchical data, but mostly at the level of the generative dynamics~\cite{sclocchi2025phase} and through the interplay between generalization and memorization of the training set in numerical experiments~\cite{favero2025bigger,garnier2026biased}. In our context, discrete diffusion on the broadcast model, e.g.\ following the approach of Ref.~\cite{elgorni2026}, instead appears as a promising setting to obtain analytical results on these questions beyond continuous diffusion of Gaussian data~\cite{biroli2024dynamical,bonnaire2025why}. It could in particular provide a solvable instance of the biased generalization of Ref.~\cite{garnier2026biased}, where it is hypothesized that bias toward training data can be caused by higher-order statistics being incorrectly resolved due to the finiteness of the dataset, a mechanism that by construction requires non-Gaussian data.

\begin{acknowledgments}
    The work of J.\,G.-B. was supported by the European Union’s Horizon Europe program under the Marie Skłodowska-Curie grant agreement No.~101210798. We thank H. Tabanelli and L. Gianmanco for useful discussions. We acknowledge the use of large language models throughout the research process, which were notably used to write numerical code and figure scripts, to help with derivations and to suggest edits to the manuscript. All results were checked by the authors, who take full responsibility for the content, and wrote the final text.
\end{acknowledgments}

\bibliography{refs}

\appendix

\section{\texorpdfstring{Belief propagation}{Belief propagation}}
\label{app:bp}

Since the generative model is a tree, the Bayes-optimal root predictor can be computed exactly with belief propagation (BP)~\cite{mezard2009information}, by passing messages upwards along the tree, from the leaves to the root. First, let us take advantage of the two-state nature of the spins to write the entries of the transition matrix $\bm\pi(\varepsilon)$ as
\begin{equation}
    \pi_{x_ix_j} = \frac{1+\theta x_i x_j}{2}, \qquad \theta = 1-2\varepsilon,
\end{equation}
where $x_i$ is a parent spin and $x_j$ one of its children. Following the standard notation of Ref.~\cite{mezard2009information}, each edge $(ij)$ is a factor node $b$, and the message sent by $b$ to the parent is updated as
\begin{equation}
    \hat{\nu}_{b\to i}(x_i) = \sum_{x_j = \pm 1} \pi_{x_ix_j}\, \nu_{j\to b}(x_j).
\end{equation}
We can reduce the probability vectors $\nu$ to a scalar magnetization,
\begin{equation}
    m_j = \nu_{j\to b}(+1) - \nu_{j\to b}(-1) = 2\nu_{j\to b}(+1) - 1,
\end{equation}
which allows the rewriting
\begin{equation}
    \nu_{j\to b}(x_j) = \frac{1 + m_j x_j}{2}.
    \label{eq:message_mag}
\end{equation}
The symbol $m$ is local to this appendix and should not be confused with the order parameters of the learning problem. Replacing in the BP update equation,
\begin{equation}
\begin{aligned}
    \hat{\nu}_{b\to i}(x_i) &= \sum_{x_j = \pm 1} \frac{1+\theta x_i x_j}{2}\, \frac{1+m_j x_j}{2}\\
    &= \frac{1 + \theta m_j x_i}{2}.
\end{aligned}
\end{equation}
Denoting by $j,k$ the two children of $i$, connected to it by the factors $b,c$, and by $a$ the edge to its own parent, the update at the variable node $i$ multiplies the incoming messages $\hat\nu_{b\to i}$ and $\hat\nu_{c\to i}$,
\begin{equation}
    \nu_{i\to a}(x_i) \propto (1+\theta m_j x_i)(1+\theta m_k x_i).
\end{equation}
Using $x_i^2=1$ and enforcing the normalization leads to
\begin{equation}
    \nu_{i\to a}(x_i) = \frac{1}{2} + \frac{1}{2}\, \frac{\theta (m_j + m_k)}{1+\theta^2 m_j m_k}\, x_i.
\end{equation}
By identification with Eq.~\eqref{eq:message_mag}, it follows that BP reduces to propagating the magnetization up the tree, from a pair of children $(j,k)$ to their parent $i$, as
\begin{equation}
    m_i = \frac{\theta(m_j + m_k)}{1+\theta^2 m_j m_k},
    \label{eq:bp-update}
\end{equation}
or equivalently, with $m=\tanh h$, $h_i = \operatorname{artanh}(\theta m_j) + \operatorname{artanh}(\theta m_k)$. The recursion is initialized at the leaves, which are observed, with $m_i = x_i$. Since the graph is a tree, BP is exact~\cite{mezard2009information}. At the root, the uniform prior gives the same update, so that $m_0 = \E{x_0\mid x_1,\dots,x_N}$. The Bayes-optimal estimator is therefore $\hat x_0^{\mathrm{BP}} = \sign m_0$, with a fair coin on ties, and it is correct with probability $(1+|m_0|)/2$ given the leaves.

This iteration is, at this stage, completely instance-specific, as $m_j$ and $m_k$ depend on the subtrees located below these nodes. Their statistics, however, obey a closed recursion in distribution. Flipping all spins of a subtree maps its magnetization $m$ to $-m$, so that, given the spin $x_i$ of its top node, $m_i$ is distributed as $x_i \tilde m$, where $\tilde m$ denotes the root-aligned magnetization, i.e., conditioned on a positive top spin. With the edge signs $\eta_1,\eta_2$ of Sec.~\ref{sec:census}, $\Pp(\eta=-1)=\varepsilon$, the root-aligned magnetization of a depth-$\ell$ tree then obeys
\begin{equation}
    \tilde m_\ell \overset{d}{=} \frac{\theta\big(\eta_1 \tilde m^{(1)}_{\ell-1} + \eta_2 \tilde m^{(2)}_{\ell-1}\big)}{1+\theta^2\eta_1\eta_2\, \tilde m^{(1)}_{\ell-1}\tilde m^{(2)}_{\ell-1}},
    \label{eq:bp-rde}
\end{equation}
with $\tilde m_0 = 1$, where $\tilde m^{(1)}_{\ell-1},\tilde m^{(2)}_{\ell-1}$ are independent copies of $\tilde m_{\ell-1}$, independent of the edge signs. This is the non-linear analogue of the census recursion, Eq.~\eqref{eq:smoothing}. By symmetry, the BP error follows from the root-aligned law alone, in parallel with the census error of Sec.~\ref{sec:census},
\begin{equation}
\begin{aligned}
    \mathcal E^{(\ell)}_{\rm BP} &= \Pp(\tilde m_\ell < 0) + \frac{1}{2} \Pp(\tilde m_\ell = 0)\\
    &= \frac{1}{2}\big(1 - \E{|\tilde m_\ell|}\big).
\end{aligned}
\label{eq:bp-error}
\end{equation}
The recursion defines the finite-depth law exactly, but enumerating its support quickly becomes impractical. We therefore approximate it by population dynamics, as for the census, whereby a pool of $M$ messages initialized at $\tilde m = 1$ is updated by replacing each member with the right-hand side of Eq.~\eqref{eq:bp-rde}, evaluated on two members drawn at random from the current pool and two fresh edge signs. After $\ell$ updates, the pool samples $\tilde m_\ell$ up to finite-$M$ fluctuations, and the $\ell\to\infty$ limit is the distributional fixed point of Eq.~\eqref{eq:bp-rde}. Since Eqs.~\eqref{eq:smoothing} and~\eqref{eq:bp-rde} share the same edge signs, both recursions can be applied jointly to a pool of pairs $(\tilde m, c)$, which samples the joint law of the BP and census outputs on the same trees.

Finally, the KS criterion is recovered as the linear stability of the trivial fixed point $\tilde m=0$ of Eq.~\eqref{eq:bp-rde}, which carries no information about the root. To linear order, $\tilde m_\ell \simeq \theta(\eta_1 \tilde m^{(1)}_{\ell-1} + \eta_2 \tilde m^{(2)}_{\ell-1})$, and averaging over $\E{\eta}=\theta$,
\begin{equation}
    \E{\tilde m_\ell} \simeq 2\theta^2\, \E{\tilde m_{\ell-1}} = r\, \E{\tilde m_{\ell-1}}.
\end{equation}
The trivial fixed point thus becomes unstable for $r = 2\theta^2 > 1$, i.e., $\varepsilon < \varepsilon_c$. Linear stability alone does not exclude a non-trivial fixed point coexisting with the trivial one for $r<1$, and this indeed happens for larger alphabets, where reconstruction is possible below KS~\cite{mezard2006reconstruction,mossel2004survey}. For the binary symmetric channel, however, it is a theorem that reconstruction is impossible for $r\le 1$~\cite{bleher1995purity,evans2000broadcasting}, so that BP and the census share the same threshold $\varepsilon_c$, and only differ by their accuracy for $\varepsilon<\varepsilon_c$.

\begin{figure}[t]
 \centering
 \includegraphics[width=\linewidth]{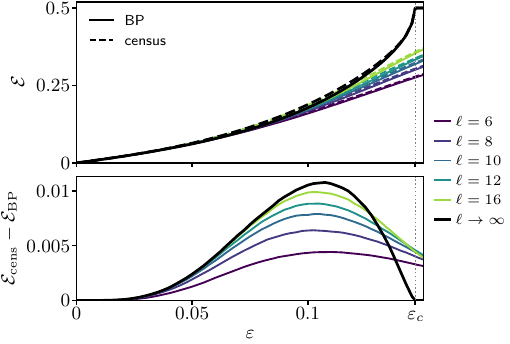}
 \caption{Bayes-optimal (BP) versus census reconstruction of the root. Top: error of BP (solid) and of the census (dashed) as a function of $\varepsilon$, for increasing depths $\ell$ and $\ell\to\infty$ (black), from a joint population dynamics of both estimators. Bottom: their difference, which remains small at all noise levels and, at infinite depth, vanishes at the KS threshold $\varepsilon_c$, where both estimators fail.}
 \label{fig:bp-census}
\end{figure}

The census and BP errors, obtained from the joint population dynamics, are compared in Fig.~\ref{fig:bp-census}. Their difference remains small at all noise levels. It notably vanishes at small $\varepsilon$, where both estimators are nearly perfect, and (at infinite depth) at the threshold where both fail, while it is largest at intermediate noise, increasing with the depth of the tree.

\section{\texorpdfstring{Exact census law}{Exact census law}}
\label{app:census-law}

By the spin-flip symmetry of the broadcast process let us fix $y=+1$ without loss of generality. Let $n_k$ be the number of $+1$ leaves of a depth-$k$ subtree whose root is $+1$, and $p_k(n)=\Pp(n_k=n)$ its law on $n=0,\dots,2^k$. A depth-$k$ subtree is the union of the two depth-$(k-1)$ subtrees rooted at the children of its root. Each child is $+1$ with probability $1-\varepsilon$, in which case its count has law $p_{k-1}$, and $-1$ with probability $\varepsilon$, in which case, by spin-flip symmetry, its count of $+1$ leaves has the flipped law $p_{k-1}(2^{k-1}-n)$. The count of a child subtree is therefore distributed as
\begin{equation}
 g_{k-1}(n)=(1-\varepsilon)\,p_{k-1}(n)+\varepsilon\,p_{k-1}(2^{k-1}-n),
 \label{eq:child-law}
\end{equation}
and since the two child subtrees are independent, $p_k$ is the discrete convolution $p_k=g_{k-1}*g_{k-1}$,
\begin{equation}
 p_k(n)=\sum_{m=0}^{2^{k-1}}g_{k-1}(m)\,g_{k-1}(n-m),
 \label{eq:node-recursion}
\end{equation}
with $g_{k-1}(m)=0$ outside $0\le m\le2^{k-1}$, starting from $p_0(n)=\delta_{n,1}$. This is the exact finite-depth counterpart of the recursion in distribution \eqref{eq:smoothing}: the edge signs $\eta_{1,2}$ are the mixture with the flip probability of \eqref{eq:child-law}, and the sum of the two subtree censuses is the convolution \eqref{eq:node-recursion}. Equivalently, one may reveal the tree level by level and propagate the number of $+1$ nodes at each level through a binomial transfer kernel, but the recursion over subtrees is simpler, as it only involves the law of a single scalar per depth.

Since $C^{(\ell)}=2n_\ell-N$, the normalized census takes the values
\begin{equation}
 \Pp\!\left(c_\ell=\frac{2n-N}{(2\theta)^\ell}\right)=p_\ell(n),
 \qquad n=0,\dots,N,
 \label{eq:census-law}
\end{equation}
on a lattice of spacing $2/(2\theta)^\ell$, which shrinks to zero for $2\theta>1$, in particular for $r>1$.

Equation~\eqref{eq:census-floor} of the main text describing the census accuracy follows from conditioning the census recursion, Eq.~\eqref{eq:smoothing}, on the two root edges. Indeed, the sign of $c_\ell$ is that of $S=\eta_1c^{(1)}_{\ell-1}+\eta_2c^{(2)}_{\ell-1}$, and with ties counted as one half the census error is $\Pp(S<0)+\tfrac12\Pp(S=0)$. If both edges are kept, which occurs with probability $(1-\varepsilon)^2$, then $S=c^{(1)}_{\ell-1}+c^{(2)}_{\ell-1}$, and we denote its error by $a_{\ell-1}$. If both edges flip, with probability $\varepsilon^2$, then $S$ changes sign and the error becomes $1-a_{\ell-1}$. If exactly one edge flips, with probability $2\varepsilon(1-\varepsilon)$, then $S=\pm(c^{(1)}_{\ell-1}-c^{(2)}_{\ell-1})$, which is symmetric about zero because the two subtree censuses are independent and identically distributed, so that the error is exactly $\tfrac12$. Summing the three cases,
\begin{equation}
\begin{aligned}
 \mathcal E^{(\ell)}_{\rm cens}&=(1-\varepsilon)^2a_{\ell-1}+\varepsilon^2(1-a_{\ell-1})+\varepsilon(1-\varepsilon)\\
 &=\varepsilon+(1-2\varepsilon)\,a_{\ell-1},
\end{aligned}
\end{equation}
which is Eq.~\eqref{eq:census-floor}, and is at least $\varepsilon$ since $a_{\ell-1}\ge0$ and $\varepsilon<1/2$.

In terms of implementation, the step $k$ of \eqref{eq:node-recursion} convolves two vectors of length $2^{k-1}+1$, so the cost is dominated by the last step and is $O(N^2)$ overall with direct summation, or $O(N\log N)$ using the fast Fourier transform. The finite-depth saddle calculations use the exact discrete recursion. Its support grows exponentially with $\ell$, whereas population dynamics keeps a fixed sample size and provides a practical route to the limiting law. The numerical implementations and the approximations used for the infinite-depth averages are described in Appendix~\ref{app:numerics}.

\section{\texorpdfstring{Moments of the limiting census}{Moments of the limiting census}}
\label{app:moments}

For $r>1$ the normalized census $c_\ell$ converges to a limit $c$ (Sec.~\ref{sec:census}), whose law solves the distributional fixed-point equation
\begin{equation}
 c\overset{d}{=}\frac{\eta_1c^{(1)}+\eta_2c^{(2)}}{2\theta},
 \label{eq:fixed-point}
\end{equation}
where $c^{(1)},c^{(2)}$ are independent copies of $c$, independent of the edge signs $\eta_1,\eta_2$. We compute its moments $M_p=\E{c^p}$ order by order, and check at the end of this appendix that they are the limits of the moments of $c_\ell$.

Raising \eqref{eq:fixed-point} to the power $p$, expanding the binomial and averaging over the independent signs and subtrees gives
\begin{equation}
 M_p=\sum_{j=0}^p\binom pj \psi_j\psi_{p-j}M_jM_{p-j},
 \label{eq:moment-rec}
\end{equation}
with $M_0=1$ and
\begin{equation}
 \psi_j=\frac{\E{\eta^j}}{(2\theta)^j}=
 \begin{cases}
  \theta/(2\theta)^j, & j\text{ odd},\\
  1/(2\theta)^j, & j\text{ even}.
 \end{cases}
\end{equation}
We first notice that $M_p$ enters the right-hand side only through the terms $j=0$ and $j=p$, each with coefficient $\psi_p$, so that the equation in fact becomes
\begin{equation}
 (1-2\psi_p)M_p=\sum_{j=1}^{p-1}\binom pj \psi_j\psi_{p-j}M_jM_{p-j},
 \label{eq:moment-tri}
\end{equation}
and the moments follow order by order whenever $1-2\psi_p\neq0$. For $p=1$, both sides vanish since $2\psi_1=1$. This reflects the invariance of \eqref{eq:fixed-point} under $c\to tc$, which leaves the mean undetermined, and we set $M_1=1$, inherited from $\E{c_\ell}=1$.

Solving \eqref{eq:moment-tri} successively, with right-hand sides $1/2$, $3M_2/(4\theta^2)$ and $M_3/(2\theta^2)+3M_2^2/(8\theta^4)$ for $p=2,3,4$, we obtain
\begin{equation}
\begin{aligned}
 M_2&=\frac{\theta^2}{2\theta^2-1}=\frac{r}{2(r-1)},\\
 M_3&=\frac{3\theta^2}{(2\theta^2-1)(4\theta^2-1)}
 =\frac{3r}{2(r-1)(2r-1)},\\
 M_4&=\frac{3\theta^4(12\theta^2-5)}{(2\theta^2-1)^2(4\theta^2-1)(8\theta^4-1)}\\
 &=\frac{3r^2(6r-5)}{4(r-1)^2(2r-1)(2r^2-1)}.
\end{aligned}
\end{equation}
The second cumulant $\kappa_2=M_2-1$ recovers $\gamma_\infty$ of \eqref{eq:gamma-inf}. The third and fourth cumulants, $\kappa_3=M_3-3M_2+2$ and $\kappa_4=M_4-4M_3-3M_2^2+12M_2-6$, read
\begin{equation}
\begin{aligned}
 \kappa_3&=-\frac{2(1-\theta^2)}{4\theta^2-1}=-\frac{2-r}{2r-1},\\
 \kappa_4&=\frac{6(1-\theta^2)(4\theta^4-\theta^2-1)}{(4\theta^2-1)(8\theta^4-1)}\\
 &=\frac{3(2-r)(2r^2-r-2)}{2(2r-1)(2r^2-1)}.
\end{aligned}
\end{equation}
Normalizing by $\gamma_\infty=(2-r)/[2(r-1)]$, the skewness and the excess kurtosis are
\begin{equation}
\begin{aligned}
 \frac{\kappa_3}{\gamma_\infty^{3/2}}
 &=-\frac{2(2\theta^2-1)^{3/2}}{(4\theta^2-1)(1-\theta^2)^{1/2}}\\
 &=-\frac{2\sqrt2\,(r-1)^{3/2}}{(2r-1)(2-r)^{1/2}},\\
 \frac{\kappa_4}{\gamma_\infty^{2}}
 &=\frac{6(r-1)^2(2r^2-r-2)}{(2-r)(2r-1)(2r^2-1)}.
\end{aligned}
\label{eq:skew-kurt}
\end{equation}
These values agree with the cumulants estimated by population dynamics, and with the asymptotic moments of the broadcast sum obtained independently in Ref.~\cite{gaitonde2026hierarchical}. The skewness is strictly negative throughout $1<r<2$, i.e.\ for every $0<\varepsilon<\varepsilon_c$, and vanishes only on approach to KS, as $-2\sqrt2\,(r-1)^{3/2}$. The excess kurtosis instead changes sign at the positive root of $2r^2-r-2$, namely $r_\kappa=(1+\sqrt{17})/4\approx1.28$, or $\varepsilon_\kappa\approx0.1$. It is positive for $\varepsilon<\varepsilon_\kappa$, negative for $\varepsilon_\kappa<\varepsilon<\varepsilon_c$, and vanishes as $-6(r-1)^2$ at KS. A test based on the kurtosis alone would thus wrongly suggest a Gaussian census near $\varepsilon\approx0.1$, where the skewness is still markedly non-zero. Near KS, the skewness also dominates the kurtosis, since $(r-1)^{3/2}\gg(r-1)^2$. This is why we use the skewness as the indicator of non-Gaussianity.

At small noise, $\theta=1-2\varepsilon$ gives
\begin{equation}
 \kappa_2\simeq4\varepsilon,\qquad
 \kappa_3\simeq-\frac{8\varepsilon}{3},\qquad
 \kappa_4\simeq\frac{16\varepsilon}{7},
\end{equation}
so that the skewness $\simeq-1/(3\sqrt\varepsilon)$ and the excess kurtosis $\simeq1/(7\varepsilon)$ diverge, a signature of rare flips near the root rather than of a weak central-limit correction, consistent with the error floor~\eqref{eq:census-floor}.

The same expansion applies at finite depth. Applied to \eqref{eq:smoothing}, it maps the moments $M_p^{(\ell-1)}=\E{c_{\ell-1}^p}$ to
\begin{equation}
 M_p^{(\ell)}=\sum_{j=0}^p\binom pj \psi_j\psi_{p-j}M_j^{(\ell-1)}M_{p-j}^{(\ell-1)},
\end{equation}
starting from $M_p^{(0)}=1$ for all $p$. This is now an explicit iteration rather than a fixed-point condition. For $p=2$ it reads $\gamma_\ell=(\gamma_{\ell-1}+1-\theta^2)/r$ with $\gamma_0=0$, whose solution is $\gamma_\ell=\gamma_\infty(1-r^{-\ell})$, as in \eqref{eq:gamma-inf}. At each order $p\ge2$, $M_p^{(\ell)}$ is an affine function of $M_p^{(\ell-1)}$ with slope $2\psi_p\le2(2\theta)^{-p}\le1/r$, since $2\theta\ge r>1$, driven by lower moments. By induction on $p$, all moments converge to the fixed-point values $M_p$, with corrections $O(r^{-\ell})$. Since $c_\ell$ is a martingale converging to $c$~\cite{kesten1966limit,biggins1977martingale}, and all its moments remain bounded, these limits are the moments of $c$.

\section{\texorpdfstring{Replica computation}{Replica computation}}
\label{app:replica}

We derive the saddle-point free energy $f^{(\ell)}$ of Eqs.~\eqref{eq:action} and~\eqref{eq:g-terms} using the celebrated replica trick $\langle\log Z_\beta\rangle=\lim_{n\to0}n^{-1}\log\langle Z_\beta^n\rangle$, evaluating $\langle Z_\beta^n\rangle$ for an integer number of replicas $n$ before continuing it to $n\to0$. The steps are standard for the perceptron, including for Gaussian-mixture inputs with a generic covariance, and we refer to Refs.~\cite{Engel_VandenBroeck_2001,loureiro2021learning} for detailed derivations. We keep them brief here and focus on what is specific to the hybrid model, the separate treatment of the census direction.

A root-aligned input can be split as
\begin{equation}
 y\bm x=s_\ell c_\ell\,\ee+\bm x_\perp ,
 \label{eq:rep-hybrid}
\end{equation}
where $\bm x_\perp$ is orthogonal to $\ee$, centered, of covariance $\bm\Sigma_\perp$ and uncorrelated with $c_\ell$. We write $\bm w^a=\sqrt N m_a\ee+\bm w^a_\perp$ for the replicas $a=1,\dots,n$ and set $\mb_a=s_\ell m_a$. The margin of replica $a$ on sample $\nu$ is
\begin{equation}
 \frac{y^\nu\bm w^a\cdot\bm x^\nu}{\sqrt N}=\mb_a c^\nu_\ell+\xi^\nu_a,\qquad
 \xi^\nu_a=\frac{\bm w^a_\perp\cdot\bm x^\nu_\perp}{\sqrt N},
\end{equation}
so that at fixed weights the data enter only through the joint law of $c^\nu_\ell$ and the transverse fields $\xi^\nu_a$. As stated in Sec.~\ref{sec:proportional}, we take these fields jointly Gaussian, centered and independent of $c^\nu_\ell$, with covariance $Q^\perp_{ab}=\bm w^{a\top}_\perp\bm\Sigma_\perp\bm w^b_\perp/N$. Since $\ee$ is an eigenvector of $\bm\Sigma$, this coincides with $\bm w^{a\top}_\perp\bm\Sigma\bm w^b_\perp/N$. Requiring this Gaussian law for all weights amounts to taking $\bm x_\perp=\bm\Sigma_\perp^{1/2}\bm z_\perp$ with $\bm z_\perp$ a standard Gaussian vector independent of $c_\ell$, leading to what we refer to as a \textit{hybrid} model. The regularization term splits as $\lambda\|\bm w^a\|^2=N\lt_\ell\mb_a^2+\lambda\|\bm w^a_\perp\|^2$. As the samples are independent, the data enter only through $\mb_a$ and $Q^\perp$, and
\begin{equation}
\begin{aligned}
 \langle Z_\beta^n\rangle&=\int\prod_a\dd\mb_a\,
 \e^{-\frac{\beta N\lt_\ell}{2}\sum_a\mb_a^2}
 \int\dd Q^\perp\,\e^{N\mathcal G(Q^\perp)}\,\mathcal Z_n^P,\\
 \mathcal Z_n&=\Big\langle\prod_a\e^{-\beta V(\mb_ac_\ell+\xi_a)}
 \Big\rangle_{c,\,\bm\xi\sim\mathcal N(0,Q^\perp)},
\end{aligned}
\label{eq:rep-Zn}
\end{equation}
up to a constant Jacobian. Here $\e^{N\mathcal G}$ is the ridge-weighted volume of transverse replicas at fixed overlaps,
\begin{multline}
 \e^{N\mathcal G(Q^\perp)}=\int\prod_a\dd\bm w^a_\perp\,
 \e^{-\frac{\beta\lambda}{2}\sum_a\|\bm w^a_\perp\|^2}\\
 \times\prod_{a\le b}\delta\!\left(NQ^\perp_{ab}
 -\bm w^{a\top}_\perp\bm\Sigma\bm w^b_\perp\right).
 \label{eq:transverse_weights}
\end{multline}
We represent the constraints by the exponential factor
\begin{equation}
 \exp\Big[-\frac{N}{2}\sum_{ab}\hat Q_{ab}Q^\perp_{ab}
 +\frac{1}{2}\sum_{ab}\hat Q_{ab}\,\bm w^{a\top}_\perp\bm\Sigma\bm w^b_\perp\Big],
 \label{eq:rep-convention}
\end{equation}
with a symmetric conjugate matrix $\hat Q$. %

We take a replica-symmetric ansatz, justified by the convexity of the loss, which in light of the $\beta \to \infty$ limit to be taken we write as
\begin{equation}
 Q^\perp_{ab}=\qp+\frac{\chi}{\beta}\delta_{ab},\quad
 \hat Q_{ab}=\beta^2\hat q-\beta\hat\chi\,\delta_{ab},\quad
 \mb_a=\mb,
\end{equation}
so that $\qp=Q^\perp_{a\neq b}$ and $\chi=\lim_{\beta\to\infty} \beta(Q^\perp_{aa}-\qp)$, and we collect the contributions to $-(\beta Nn)^{-1}\log\langle Z_\beta^n\rangle$ as $n\to0$ and then $\beta\to\infty$. The first term of \eqref{eq:rep-convention} contributes $(\hat q\chi-\hat\chi\qp)/2=g_I$, and the ridge on the signal coordinate contributes $\lt_\ell\mb^2/2$.

In the weight integral, the second term of \eqref{eq:rep-convention} is $\tfrac{\beta^2\hat q}{2}\|\bm\Sigma^{1/2}\sum_a\bm w^a_\perp\|^2 -\tfrac{\beta\hat\chi}{2}\sum_a\bm w^{a\top}_\perp\bm\Sigma\bm w^a_\perp$. A Hubbard--Stratonovich transformation with $\bm u\sim\mathcal N(0,I_N)$ linearizes the first part and decouples the replicas. Each one then has a Gaussian integral on the transverse subspace, with precision $\beta(\lambda+\hat\chi\bm\Sigma_\perp)$ and source $\beta\sqrt{\hat q}\,\bm\Sigma_\perp^{1/2}\bm u$. Completing the square and averaging over $\bm u$ at $n\to0$, the contribution from the integral over the transverse weights Eq.~\eqref{eq:transverse_weights} finally gives
\begin{equation}
 \frac{\beta\hat q}{2}\Tr\frac{\bm\Sigma_\perp}{\lambda+\hat\chi\bm\Sigma_\perp}
 =\frac{\beta N\hat q}{2}\Big\langle\frac{\mu}{\lambda+\mu\hat\chi}
 \Big\rangle_{\rho_{\perp,\ell}},
\end{equation}
the trace running over the transverse subspace, where the spectrum of $\bm\Sigma_\perp$ is that of $\bm\Sigma$ with one copy of $\ms$ removed, i.e.\ $\rho_{\perp,\ell}$ of \eqref{eq:measures} after normalization by $N$. The transverse ridge is not rescaled, so the bare $\lambda$ appears. With the signal ridge, this builds $g_S$. The log-determinant is $O(N\log\beta)$ and drops out after division by $\beta$.

In the data term, the replica-symmetric fields are $\xi_a=\sqrt{\qp}\,\zeta+\sqrt{\chi/\beta}\,\zeta_a$, with $\zeta,\zeta_a$ independent standard Gaussians, and writing $h=\mb c_\ell+\sqrt{\qp}\,\zeta$,
\begin{equation}
 \mathcal Z_n=\Big\langle\Big(\big\langle
 \e^{-\beta V(h+\sqrt{\chi/\beta}\,\zeta_a)}\big\rangle_{\zeta_a}\Big)^n\Big\rangle_{c,\zeta}.
\end{equation}
To first order in $n$, $\log\mathcal Z_n=n\langle\log\langle\cdots\rangle_{\zeta_a}\rangle_{c,\zeta}+O(n^2)$. Changing variables to $z=h+\sqrt{\chi/\beta}\,\zeta_a$,
\begin{equation}
 \big\langle\e^{-\beta V(h+\sqrt{\chi/\beta}\,\zeta_a)}\big\rangle_{\zeta_a}
 =\sqrt{\frac{\beta}{2\pi\chi}}\int\dd z\,\e^{-\beta\left[\frac{(z-h)^2}{2\chi}+V(z)\right]},
\end{equation}
and since the bracket has a unique minimum for convex $V$, Laplace's method gives exactly $-\beta^{-1}\log\langle\cdots\rangle_{\zeta_a}\to\min_z[(z-h)^2/(2\chi)+V(z)]$ as $\beta\to\infty$, the prefactor contributing only $O(\beta^{-1}\log\beta)$. The data thus contribute $\alpha g_E$ of the main text. The census enters only through this one-dimensional average over its exact law. Summing the contributions and extremizing at large $N$ gives $f^{(\ell)}$.

Stationarity of~\eqref{eq:action} with respect to $\hat q$, $\chi$, $\hat\chi$, $\qp$ and $\mb$ gives, in this order, the five equations~\eqref{eq:saddle}. The derivatives of $g_E$ follow from $z_\star=h-\chi V'(z_\star)$, which gives $V'(z_\star)$ for the derivative of the minimum in $h$ and $-V'(z_\star)^2/2$ in $\chi$, together with Gaussian integration by parts in $\zeta$ at fixed $c_\ell$ and $\partial_hz_\star=1/(1+\chi V''(z_\star))$ for the derivative in $\qp$. A fresh sample has margin $h$ itself, which gives the test error \eqref{eq:test-error}.

The Gaussian saddle may be recovered by taking $c_\ell \sim \mathcal{N}(1,\gamma_\ell)$. Stein's lemma then gives
\begin{equation}
 \langle c_\ell V'(z_\star)\rangle_{c,\zeta}
 =\langle V'(z_\star)\rangle_{c,\zeta}
 +\gamma_\ell\mb\Big\langle\frac{V''(z_\star)}{1+\chi V''(z_\star)}
 \Big\rangle_{c,\zeta},
\end{equation}
and with the equation for $\hat\chi$ the signal equation becomes
\begin{equation}
 (\lt_\ell+\gamma_\ell\hat\chi)\,\mb=-\alpha\langle V'(z_\star)\rangle_{c,\zeta}.
 \label{eq:rep-stein}
\end{equation}
Multiplying by $s_\ell$ and using $\mb=s_\ell m$, $\lt_\ell s_\ell^2=\lambda$ and $\gamma_\ell s_\ell^2=\ms$, this is $(\lambda+\ms\hat\chi)m=-\alpha s_\ell\langle V'(z_\star)\rangle_{c,\zeta}$, the signal equation for Gaussian covariates with covariance $\bm\Sigma$. The remaining equations coincide with those of Gaussian covariates, once their overlap is identified as $q=\qp+\ms m^2$, since the full and transverse spectra are the same up to $O(1/N)$ (the top eigenvalue weight).

\section{\texorpdfstring{Spectral truncation and finite-depth corrections}{Spectral truncation and finite-depth corrections}}
\label{app:truncation}

At $\ell=\infty$ the spectral averages of Eq.~\eqref{eq:saddle} run over the infinite measure $\rho$ of \eqref{eq:rho-limit}. We truncate it and bound the resulting error, assuming $\hat\chi>0$. The integrands $\mu^j/(\lambda+\mu\hat\chi)^j$, $j=1,2$, increase with $\mu$ and saturate,
\begin{equation}
 0\le\frac{\mu}{\lambda+\mu\hat\chi}<\frac{1}{\hat\chi},\qquad 0\le\frac{\mu^2}{(\lambda+\mu\hat\chi)^2}<\frac{1}{\hat\chi^2}.
\end{equation}
Let $\rho^{(K)}=\sum_{k\le K}2^{-k}\delta_{\mu_k}$ be $\rho$ truncated at height $K$. Since the discarded mass is $\sum_{k>K}2^{-k}=2^{-K}$,
\begin{equation}
 0\le\Big\langle\frac{\mu^j}{(\lambda+\mu\hat\chi)^j}\Big\rangle_{\rho} -\Big\langle\frac{\mu^j}{(\lambda+\mu\hat\chi)^j}\Big\rangle_{\rho^{(K)}} \le\frac{2^{-K}}{\hat\chi^{j}}.
 \label{eq:trunc-spectral}
\end{equation}
By \eqref{eq:measures}, $\rho_{\perp,\ell}=\sum_{k<\ell}2^{-k}\delta_{\mu_k}+2^{-\ell}\delta_{\mu_\ell} =\rho^{(\ell)}$, meaning that the finite-depth transverse measure is exactly $\rho$ truncated at height $\ell$, and the bound \eqref{eq:trunc-spectral} holds for $\rho_{\perp,\ell}$ with $2^{-K}$ replaced by $2^{-\ell}=1/N$. The full measure $\rho_\ell$ instead moves the mass $2^{-\ell}$ of $\{\mu_k\}_{k>\ell}$ down to $\mu_\ell$, and monotonicity of the integrands gives the same bound. The discarded tail can also be replaced by its saturated value $2^{-K}\hat\chi^{-j}$. Since $1-(1+x)^{-j}\le jx$ with $x=\lambda/(\mu\hat\chi)$, the error then drops to $2^{-K}\hat\chi^{-j}\,j\lambda/(\hat\chi\mu_{K+1})=O((2r)^{-K})$ for $1<r<2$, at fixed $\lambda$ and $\hat\chi>0$.

\begin{figure}[t]
 \centering
 \includegraphics[width=\linewidth]{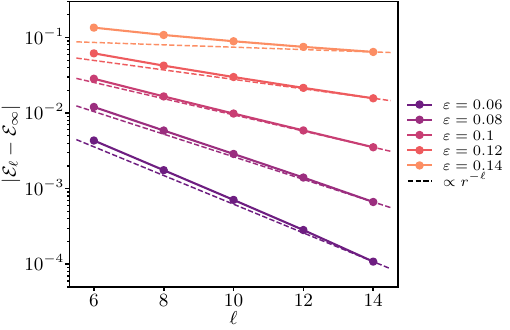}
 \caption{Distance of the finite-depth saddle error to its infinite-depth limit at $\lambda=0.1$ and $\alpha=8$, for several values of $\varepsilon$, with the census rate $r^{-\ell}$ shown dashed. Convergence slows as $\varepsilon$ approaches $\varepsilon_c$, where the depth scale $1/\log r$ diverges.}
 \label{fig:finite-depth-convergence}
\end{figure}

Figure~\ref{fig:finite-depth-convergence} shows the convergence of the finite-depth saddle error to its infinite-depth limit, at the rate $r^{-\ell}$ of the census correction, which dominates the $O(1/N)$ spectral one.

\section{Numerical protocol}
\label{app:numerics}

The empirical risk minimization points in the main text plots are obtained by logistic regression with the L-BFGS solver of scikit-learn~\cite{pedregosa2011scikit} (no intercept, inputs divided by $\sqrt N$, and the regularization convention of Eq.~\eqref{eq:erm}, which corresponds to $C=1/\lambda$ in this implementation). Each point is the test error averaged over several independent repeats, each with its own training set and its own test set of $2\times10^5$ fresh trees. The plotted intervals are approximate $95\%$ confidence intervals, given by $1.96$ times the standard error of the mean over repeats, which therefore includes both the training and the test fluctuations.

When all training labels are equal, which happens with probability $2^{1-P}$ for balanced labels, the ridge-regularized objective remains well defined and has a unique minimizer. We fit it by including every training example both as $(\bm x,y)$ and as $(-\bm x,-y)$ with weight $1/2$, which leaves every signed margin, hence the objective, unchanged, while providing the two classes required by the fitting interface.

The saddle-point equations are solved with a Julia solver, available upon request, which uses the exact census law at finite depth (Appendix~\ref{app:census-law}), population dynamics at infinite depth, and the spectral truncation of Appendix~\ref{app:truncation}.

\begin{figure}[t]
 \centering
 \includegraphics[width=\linewidth]{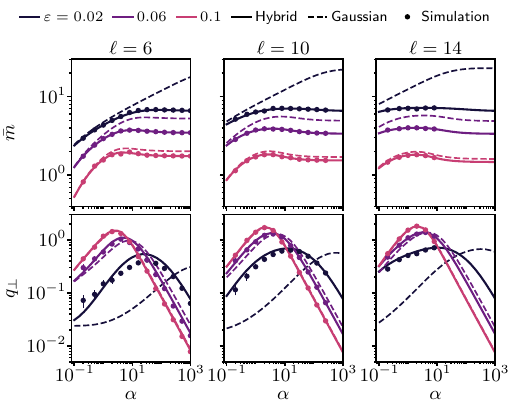}
 \caption{Order parameters of logistic regression on broadcast trees at $\lambda=0.1$, versus $\alpha$ at depths $\ell=6$, $10$ and $14$ and for three values of $\varepsilon$. Signal overlap $\mb$ (top) and transverse overlap $\qp$ (bottom), from the finite-depth hybrid saddle (solid lines), the Gaussian saddle (dashed lines) and the weights trained in the simulations of Fig.~\ref{fig:finite-depth} (dots, approximate $95\%$ intervals).}
 \label{fig:orderparams}
\end{figure}

\section{Order parameters of the trained classifiers}
\label{app:orderparams}

Given the trained weights $\bm w$ from ERM numerical experiments, the order parameters $\mb=s_\ell\,\1\cdot\bm w/N$ and $\qp=\bm w_\perp^\top\bm\Sigma_\perp\bm w_\perp/N$ follow exactly from the covariance~\eqref{eq:covariance}. Figure~\ref{fig:orderparams} compares them with the finite-depth hybrid and Gaussian saddles along the learning curves of the bottom row of Fig.~\ref{fig:finite-depth}. The hybrid saddle describes $\mb$ accurately, whereas the Gaussian saddle overestimates it, increasingly so at small $\varepsilon$ and large $\alpha$. The non-Gaussian census thus also changes the learned weights. The transverse overlap is also well described, except at the smallest noise level, where it deviates from the saddle at small depth and approaches it as the depth grows. Training on synthetic data, with the exact census law and an independent Gaussian transverse field, appears to remove the discrepancy, suggesting that it entirely stems from the dependence between census and transverse field neglected by the theory. It has little effect on the test error however. Indeed, at large signal-to-noise ratio $\mb^2/\qp$, the margin $h$ has the sign of $c_\ell$ except for the rare samples with $|c_\ell|\lesssim\sqrt{\qp}/\mb$, so that $\mathcal E\simeq\mathcal E^{(\ell)}_{\rm cens}$ regardless of the precise value of $\qp$, as shown by the perfect match with numerical simulations in the small $\varepsilon$ region of Fig.~\ref{fig:finite-depth}.

\end{document}